\documentclass[reprint,aps,prl]{revtex4-2}

\usepackage{graphicx}
\usepackage{amsmath,amssymb,bm}
\usepackage{hyperref}
\usepackage{color}

\begin{document}
\title{
Footprints of loop extrusion in  statistics of intra-chromosomal   distances: an analytically solvable model
}
\date\today

\author{Sergey Belan}
\email{sergb27@yandex.ru}
\affiliation{Landau Institute for Theoretical Physics, Russian Academy of Sciences, 1-A Akademika Semenova av., 142432 Chernogolovka, Russia}
\affiliation{National Research University Higher School of Economics, Faculty of Physics, Myasnitskaya 20, 101000 Moscow, Russia}

\author{Vladimir Parfenyev}
\affiliation{Landau Institute for Theoretical Physics, Russian Academy of Sciences, 1-A Akademika Semenova av., 142432 Chernogolovka, Russia}
\affiliation{National Research University Higher School of Economics, Faculty of Physics, Myasnitskaya 20, 101000 Moscow, Russia}

\begin{abstract} 
Active loop extrusion -- the process of formation of dynamically growing chromatin loops due to the motor activity of DNA-binding protein complexes -- is firmly established  mechanism  responsible for chromatin spatial organization at different stages of cell cycle in eukaryotes and bacteria. 
The theoretical insight into the effect of  loop extrusion on the experimentally measured statistics of chromatin conformation can be gained with an appropriately chosen polymer model.
Here we consider the simplest analytically solvable model of  interphase chromosome which is treated as ideal chain  with  disorder of sufficiently sparse random loops whose  conformations are sampled from the  equilibrium ensemble.
This framework allows us to arrive at the closed-form analytical expression for the mean-squared distance between pairs of genomic loci which is valid beyond the one-loop approximation in diagrammatic representation.
Besides, we analyse the loops-induced deviation of chain conformations from the Gaussian statistics by calculating kurtosis of probability density of the pairwise separation vector.
The presented results suggest the possible ways of estimating the characteristics of the loop extrusion process based on the experimental data on the scale-dependent statistics of intra-chromosomal pair-wise distances.
\end{abstract}

\maketitle

\begin{figure}[t]
\includegraphics[width=0.8\linewidth]{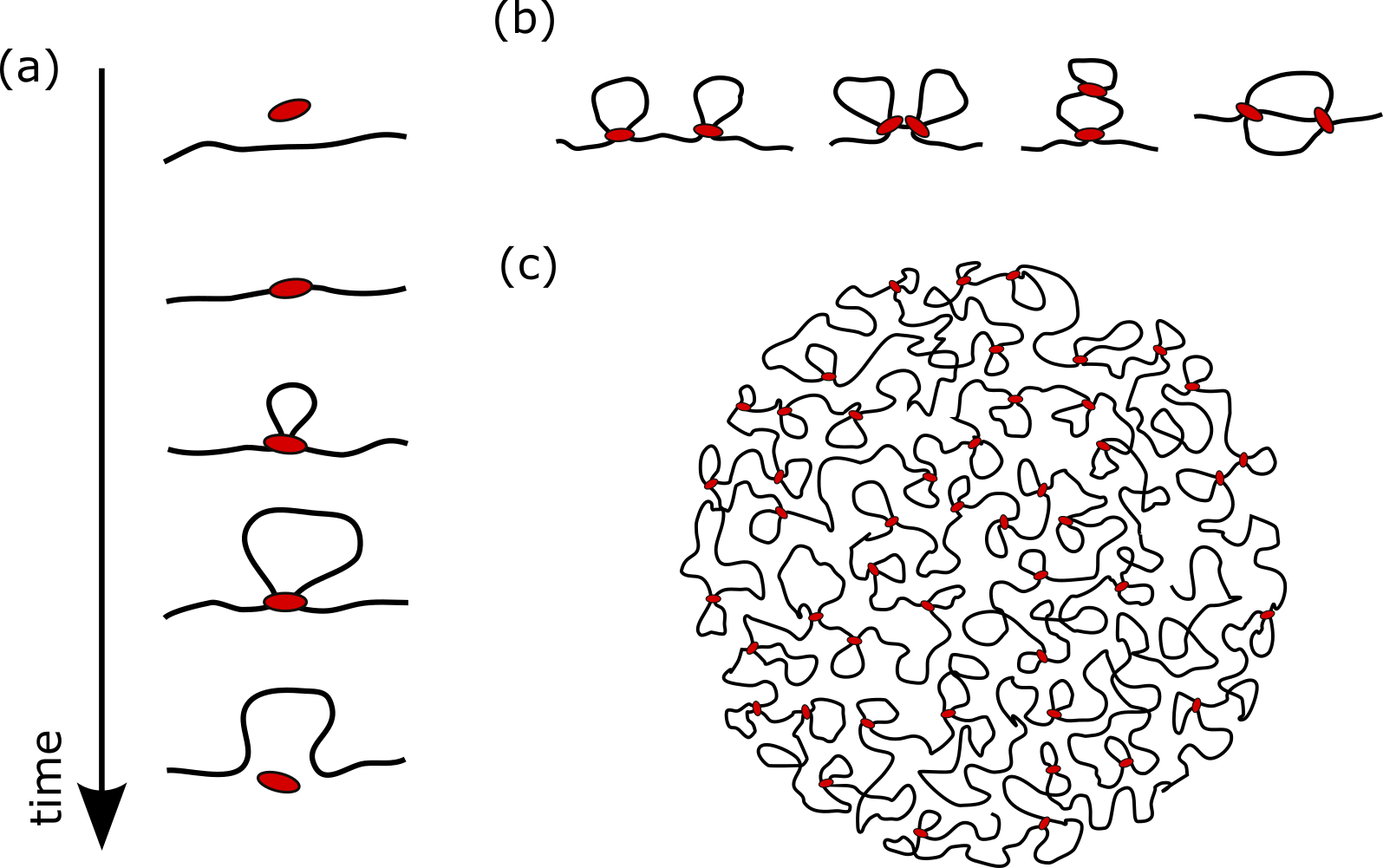}
\caption{(a) A schematic of the loop extrusion model: over time a motor protein (depicted in red) binds chromatin, extrudes a loop, and unbinds. (b) Variants of mutual arrangement of two neighboring cohesin-anchored loops. From left to right: two loops separated a gap; blocking configuration; nested configuration; two cohesins bypassing each other form a Z-loop. In sufficiently low cohesion concentration one can neglect the second and the third scenarios. (c) Polymer chain with an array of sparse random loops as a model of interphase chromosome (loop bases are depicted in red).}
\label{pic:fig1}
\end{figure}

\textit{Introduction.} 
A series of recent single-molecule experiments have shown that the structural maintenance of chromosomes proteins, such as condensin and cohesin, when binding to DNA can exhibit ATP-dependent motor activity leading to progressive growth of DNA loops \cite{ganji2018real,golfier2020cohesin,kong2020human,davidson2019dna,kim2019human,ryu2020condensin,banigan2020loop}.
These works provided long-awaited direct evidence of active loop extrusion -- a hypothetical molecular mechanism previously  introduced to explain a broad range of data on spatial organization of genome throughout the cell cycle \cite{kimura199913s,nasmyth2001disseminating,riggs1990dna}.
Incorporation of loop extrusion mechanism into polymer models of chromatin folding has proven to be successful in explaining the experimental data on three dimensional genome organization in live cells available due to explosion of super-resolution imaging methods and sequencing-based techniques.
In particular, the molecular dynamics simulation of chromatin folding accounting for the motor units that randomly bind to chromatin fiber and extrude chromatin loops until stochastically dissociating (see Fig. \ref{pic:fig1}a) 
allows to reproduce the interphase domains  observed in the  population-averaged
Hi-C maps  \cite{sanborn2015chromatin, fudenberg2016formation, 
fudenberg2017emerging, mirny2019two,banigan2020chromosome}.
Besides, computational models indicate that loop extrusion can explain condensin-mediated mitotic chromosome compaction and segregation \cite{alipour2012self, gibcus2018pathway, goloborodko2016compaction, goloborodko2016chromosome}.
Taken together, these results pave the way towards a better  understanding of  how 3d chromatin architecture regulates the genome function~\cite{hafner2022spatial}.

The growing body of experimental data calls for development of analytical models that would give easily interpretable predictions concerning effect of loop extrusion machinery on statistics of chromatin conformation avoiding the need to perform computationally intensive simulations.
Recent theoretical work \cite{polovnikov2022fractal} has shown the promise of the fractal polymer model with quenched disorder of random loops for systematization of the experimentally available statistical information on the pairwise contacts in interphase genome of higher eukaryotes for genomic scales up to several megabases. 
Here we exploit the minimalistic version of this model where chromatin is treated as ideal chain with loops disorder to describe the expected footprints  of cohesin-driven loop extrusion  in the statistics of the physical distances between pairs of genomic loci in interphase chromosome, which can potentially be extracted via state-of-art  microscopy-based techniques \cite{cattoni2017single,
ou2017chromemt,bintu2018super,nir2018walking,boettiger2020advances,kempfer2020methods, su2020genome,liu2020multiplexed, xie2021single,li2021nanoscale,gabriele2022dynamics}.

\textit{Model formulation.} 
Let us list key assumptions underlying our theoretical analysis.
First of all, based on estimates presented in previous studies \cite{polovnikov2022fractal,Goloborodko_2016}, we will assume that   for interphase chromatin the fraction of nested, blocking and Z-like loop configurations (see  Fig. \ref{pic:fig1}b) is relatively small, so that most of the  cohesin-mediated loops are separated  from each other by loops-free gaps as shown in Fig \ref{pic:fig1}c.
Since both cohesin-chromatin binding kinetics and ATP-consuming motor activity of cohesin are inherently stochastic, the array of  cohesion-mediated loops should be characterized statistically.
Given the previous assumption of a fairly low concentration of cohesin, one can treat the lengths of loops and of inter-loops gaps as statistically independent. 
Assuming additionally a constant extrusion speed, Poisson kinetics of cohesin binding/dissociation, uniform distribution of cohesin binding sites and neglecting distinct loop extrusion barriers (see, e.g., Refs.~\cite{fudenberg2017emerging,banigan2022motors}), we adopt the exponential probability densities   
for random lengths of loops and gaps with parameters $\lambda$ and $g$ denoting the mean loop length and mean gap length, respectively.  
The dimensionless ratio $\lambda/g$
 is less than or of order of unity in interphase \cite{polovnikov2022fractal,Goloborodko_2016}.
Next,   
simple estimates show that the characteristic time required for the cohesin complex to extrude a chromatin loop corresponding to a DNA region of $\sim100$ kbp, which corresponds to typical loop length in interphase estimated from in vivo Hi-C data, is long compared to the  relaxation time of such a loop \cite{polovnikov2022fractal}.
Given this argument, in our analytical calculations we will treat the loops disorder as frozen.
Finally, completely neglecting steric effects and affinity interactions, we will assume that chromatin is an ideal phantom chain with the Kuhn segment $l_{\text{eff}}$ \cite{GKh_1994}.

Summarizing the above assumptions, we arrive at a model of an equilibrium ideal chain with quenched disorder of random loops, characterized by exponential probability densities of statistically independent contour lengths of loops and gaps.
As shown in Ref.~\cite{polovnikov2022fractal}, the semi-analytical calculations and asymptotic one-loop analysis based on this model qualitatively reproduce specific shape of experimental contact probability curves universal among mammalian cells.
Also, in the work \cite{belan2022influence} this model has been used to extract one-loop predictions regarding the scale-dependent conditional probabilities of triple contacts, which can be measured with the experimental techniques for detecting multiple contacts between more than two chromatin regions \cite{darrow2016deletion,olivares2016capturing, beagrie2017complex, bintu2018super, quinodoz2018higher, allahyar2018enhancer,oudelaar2018single,ulahannan2019nanopore, kempfer2020methods, vermeulen2020multi,tavares2020multi,quinodoz2022sprite}.
In this paper, we focus on the  statistics of the physical distances between pairs of genome regions rather than on pairwise contact frequencies.
Overcoming the methodological shortcomings of the semi-analytical and perturbative approaches used in Refs. \cite{banigan2020chromosome,polovnikov2022fractal,belan2022influence}, here we present a method 
for exact summation of a diagrammatic series  which allows us to derive an analytical answer for mean-squared distance between pair of loci and  can potentially be generalized to the statistical moments of arbitrary order.
In what follows,  the key steps of derivation are outlined, whereas the technical details can be found in Appendix.

\begin{figure}[t]
\includegraphics[width=0.9\linewidth]{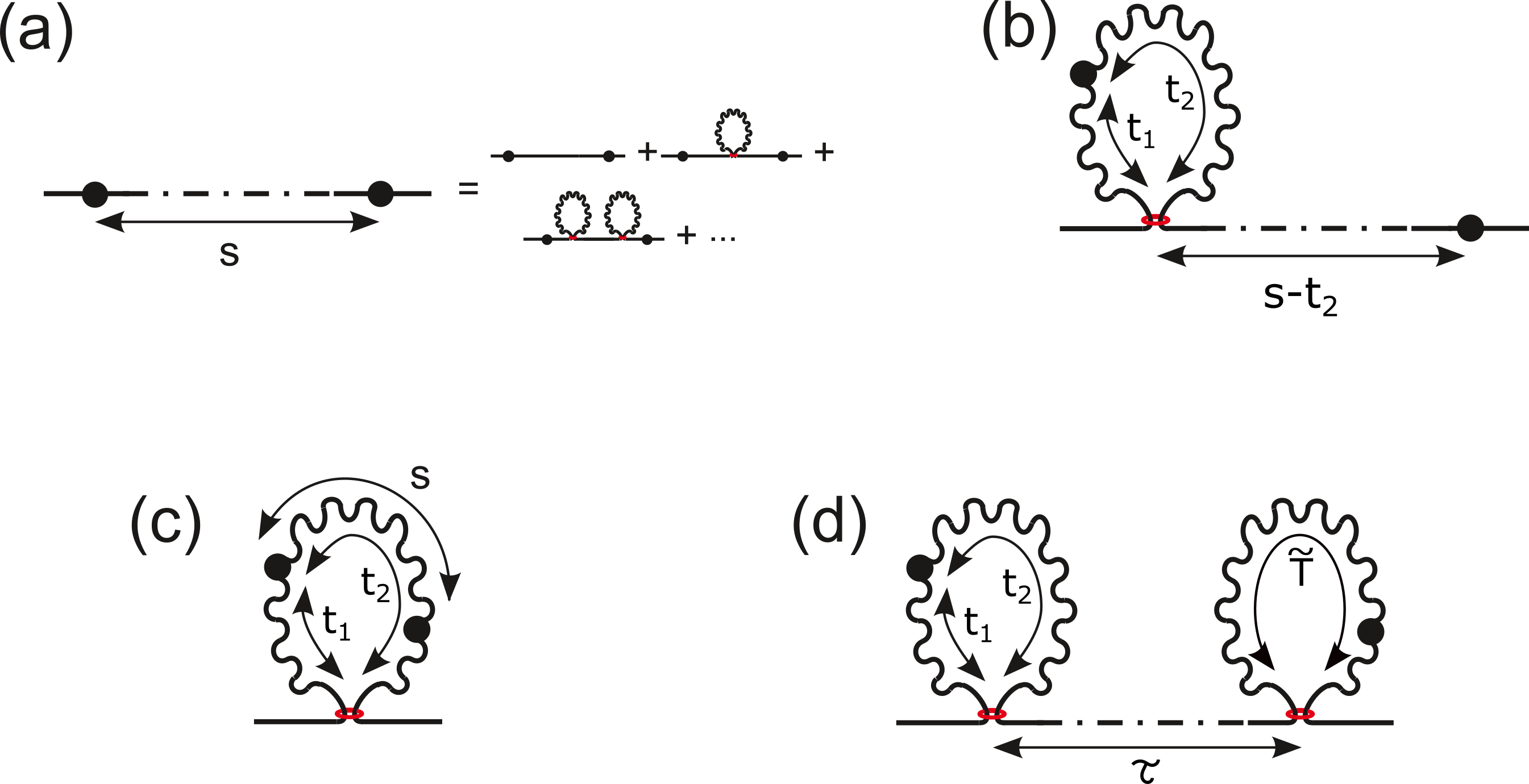}
\caption{Four classes of diagrams contributing to the MSD between two points of the ideal chain with disorder of random loops:   (a) both points  reside beyond the cohesin-mediated loops; (b) one point resides at a loop, while another point is in  inter-loop gap ($t_1\ge 0$, $0\le t_2\le s$);  (c) both points reside in the same loop ($t_1 \ge 0$, $0\le s \le t_2$); (d) the points  belong to two different loops ($t_1\ge 0$, $0\le t_2\le s$, $0\le\tau\le s-t_2$,  $\tilde{T}\ge s-t_2-\tau$). 
Note that the dash-dotted lines in diagrams (a), (b) and (d) may contain arbitrary number of random loops.}
\label{fig:diagrams}
\end{figure} 

\textit{Outline of calculations.} 
Let us denote as  $\vec R(s)$ the vector between two points of the chain separated by the contour distance $s$.
The main metric of interest for us is the  mean-squared  distance (MSD) defined as $\langle R^2(s)\rangle$, where angular brackets denote averaging of the statistics of thermal noise and random loops.
Clearly, there are four scenarios for the relative arrangement of the selected points and bases of the cohesin-mediated loops, see Fig.~\ref{fig:diagrams}.
Given this, the average physical separation can be represented as
\begin{equation}\label{eq:Prob}
\langle R^2(s) \rangle=\sum_{\alpha=a,b,c,d}\langle {\cal R}^2_{\alpha}(s| \{ A\}_{\alpha})\rangle_{\text{loops}},
\end{equation} 
where $\alpha$ enumerates the diagrams according to Fig.~\ref{fig:diagrams}, $ {\cal R}^2_{\alpha}(s| \{ A\}_{\alpha})$ is the conditional MSD obtained by averaging of $R^2(s)$ over thermal noise at fixed pattern of random loops, $\{ A\}_{\alpha}$ represents the set of random variables parametrising the corresponding diagram, and $\langle ...\rangle_{\text{loops}}$ denotes averaging over variables $\{ A\}_{\alpha}$.
Note that since the loop disorder is quenched by assumption, the averaging over thermal fluctuations precedes the averaging over the statistics of random loops in Eq.~(\ref{eq:Prob}).

 
To arrive at the MSD, one first needs to derive the conditional  expressions $ {\cal R}^2_{\alpha}(s| \{ A\}_{\alpha})$ associated with the different diagrams, depicted in Fig. \ref{fig:diagrams}. 
By virtue of the central limit theorem, the large-scale conformational statistics of the loop-free ideal chain is equivalent to that of the Brownian particle trajectory,  with time measured in the units of the polymer contour length and diffusion coefficient 
 $D = l_{\text{eff}}/6$ (see, e.g., Ref.~\cite{GKh_1994}). 
Thus, if $\lambda,g\gg l_{\text{eff}}$ and we are interested at scales $s\gg l_{\text{eff}}$, then
chromatin conformation can be thought of as alternating free Brownian paths and Brownian bridges.

In the absence of random loops, the MSD between two sites of an equilibrium Gaussian chain behaves as  ${\cal R}^2(s)=l_{\text{eff}} s$.
As follows from analysis presented in Refs.~\cite{banigan2020chromosome,polovnikov2022fractal},  the conditional MSD ${\cal R}^2_{\alpha}(s| \{ A\}_{\alpha})$ associated with fixed configuration of random loops obeys the same linear scaling law, but with an effective contour separation $\tilde s_{\alpha}[s,\{ A\}_{\alpha}]$  substituted for $s$.
More specifically, one obtains (see Appendix for details)
\begin{equation}
\label{conditional_msd_via_s_eff}
{\cal R}^2_{\alpha}(s| \{ A\}_{\alpha})= l_{\text{eff}} \tilde{s}_{\alpha}[s,\{ A\}_\alpha],
\end{equation}
where 
\begin{eqnarray}
\label{s_eff_a}
&\tilde{s}_{a}[s,x_s]=(1-x_s)s,&\\
\label{s_eff_b}
&\tilde{s}_{b}[s,t_1,t_2,x_{s-t_2}]=(1-x_{s-t_2})(s-t_2)+\frac{t_1t_2}{t_1+t_2},&\\
\label{s_eff_c}
&\tilde{s}_{c}[s,t_1,t_2]=\left(1-\frac{s}{t_1+t_2}\right)s,&\\
\label{s_eff_d}
&\tilde{s}_{d}[s,t_1,t_2,\tau,x_{\tau},\tilde{T}]=\frac{t_1t_2}{t_1+t_2}+(1-x_\tau)\tau+\frac{\tilde{t}_1\tilde{t}_2}{\tilde{t}_1+\tilde{t}_2},&
\end{eqnarray}
and $\tilde{t}_1=\tilde{T}+\tau+t_2-s$, $\tilde{t}_2=s-\tau-t_2$.
Here $t_1$, $t_2$, $\tilde T$ and $\tau$ represent the contour lengths of the segments depicted in Figs.~\ref{fig:diagrams}a, b, and d, while $x_s$, $x_{s-t_2}$ and $x_\tau$ are the fractions of contour length occupied by loops in the segments depicted by dotted lines in diagrams (a), (b) and (d), respectively.
Note, that this variables obey the  constraint $0\le x_s,x_{s-t_2},x_\tau< 1$.


Next, we should average conditional MSD $ {\cal R}^2_{\alpha}(s| \{ A\}_{\alpha})$  over the statistics of random variables $\{ A\}_{\alpha}$. 
In order to derive the corresponding statistical weights, it is convenient to introduce a two-state Markov jump process in continuous time
where time intervals are measured in the units of the polymer contour length and stochastic transitions between two states occur with the rates $\alpha_l=\lambda^{-1}$ and $\alpha_g=g^{-1}$.
Clearly, the statistics of alternating loops and gaps in our original problem are equivalent to the statistics of time intervals that this auxiliary Markov process spends in different states in the course of its stochastic dynamics.
As shown in Ref.~\cite{polovnikov2022fractal} (see also Appendix), the exact analytical expressions for statistical weights  ${\cal W}_\alpha(\{A\}_{\alpha};s)$ can be derived  from the basic properties of two-state Markov chain:
\begin{eqnarray}
\label{eq:Wa}
&{\cal W}_{a}=p_g\pi_{g\to g}(s){\cal F}(x_s),&\\
\label{eq:Wb}
&{\cal W}_{b}=2p_l\alpha^2_l e^{-\alpha_l(t_1+t_2)}\pi_{g\to g}(s-t_2) {\cal F}(x_{s-t_2}),&\\
\label{eq:Wc}
&{\cal W}_{c}=p_l\alpha_l^2e^{-\alpha_l(t_1+t_2)},&\\
\label{eq:Wd}
&{\cal W}_{d}=p_l\alpha_l^3e^{-\alpha_l(t_1+t_2+\tilde T)}\alpha_g \pi_{g\to g}(\tau) {\cal F}(x_\tau),&
\end{eqnarray} 
where $p_g=\frac{\alpha_l}{\alpha_g+\alpha_l}$ and $p_l=\frac{\alpha_g}{\alpha_g+\alpha_l}$ give the probabilities that a starting point of the  walker's trajectory belongs to a free Brownian path and loop, respectively, $\pi_{g\to g}(s)=\frac{1}{\alpha_g+\alpha_l}(\alpha_l+\alpha_ge^{-(\alpha_l+\alpha_g)s})$ is the probability to find Markov process in the gap state after time $s$ given that initially it was in the same state, and ${\cal F}(x_s)$ represents the probability density of $x_s$.
Exact expression for ${\cal F}(x_s)$ can be extracted from the Pendler's work~\cite{Pedler_1971} on the occupation time statistics of two-state Markov process and is given by Eq. (29) in Appendix.

The loop-averaged conditional MSD $\langle {\cal R}^2_{\alpha}(s| \{ A\}_{\alpha})\rangle_{\text{loops}}$ entering Eq. (\ref{eq:Prob}), is given by integration of $ {\cal R}^2_{\alpha}(s| \{ A\}_{\alpha})$  with weight  ${\cal W}_\alpha(\{A\}_{\alpha};s)$  over the variables $\{ A\}_{\alpha}$.
The main technical difficulties are associated with averaging over the random variables $x_s$, which parametrizes the expressions (\ref{s_eff_a}), (\ref{s_eff_b}) and (\ref{s_eff_d}).
An exact probability density ${\cal F}(x_s)$, while efficient for numerical analysis, is inconvenient for analytical calculations.
Note, however, the conditional MSDs defined by expression (\ref{conditional_msd_via_s_eff}) are linear with respect to the variable $x_s$. Exploiting properties of Markov bridge statistics  one obtains (see Appendix) 
\begin{eqnarray}
\label{x_s}
\langle x_s\rangle= \dfrac{1}{s} \int_0^s dt \dfrac{\pi_{g\to l}(t) \pi_{l \to g}(s-t)}{\pi_{g\to g}(s)},
\end{eqnarray}
where $\pi_{g\to l}(s)=\frac{\alpha_g}{\alpha_g+\alpha_l}(1-e^{-(\alpha_g+\alpha_l)s})$ and  $\pi_{l\to g}(s)=\frac{\alpha_l}{\alpha_g+\alpha_l}(1-e^{-(\alpha_g+\alpha_l)s})$. 
With Eq. (\ref{x_s}) we can express conditional MSDs without usage of the cumbersome formula for  ${\cal F}(x_s)$. Note that such a trick does not work in more sophisticated case of contact probability calculations where associated diagram contributions  are non-linear in $\langle x_s\rangle$ and should be analysed numerically \cite{polovnikov2022fractal,belan2022influence}.
\textit{Results.} 
Rather laborious calculation procedure finally leads us to surprisingly elegant analytical expression for the MSD
\begin{equation}\label{msd}
\langle R^2(s)\rangle= \dfrac{l_{\text{eff}}s}{1+\lambda/g} \left[ 1 + \dfrac{\lambda}{g} f_{\text{MSD}}\left( \dfrac{s}{\lambda} \right) \right], 
\end{equation}
where $f_{\text{MSD}}(z) = \frac{2}{3} (z^{-1}(1-e^{-z}) + {\cal E}_3(z))$ and ${\cal E}_n(z)=\int_{1}^{+\infty}x^{-n}e^{-zx}dx$ is the exponential integral function. 
Importantly, this result is non-perturbative in the sense that it takes into account all zoo of diagrams in our model and, thus, is formally valid for any  value of the dimensionless ratio $\lambda/g$.

\begin{figure}
\includegraphics[width=0.95\linewidth]{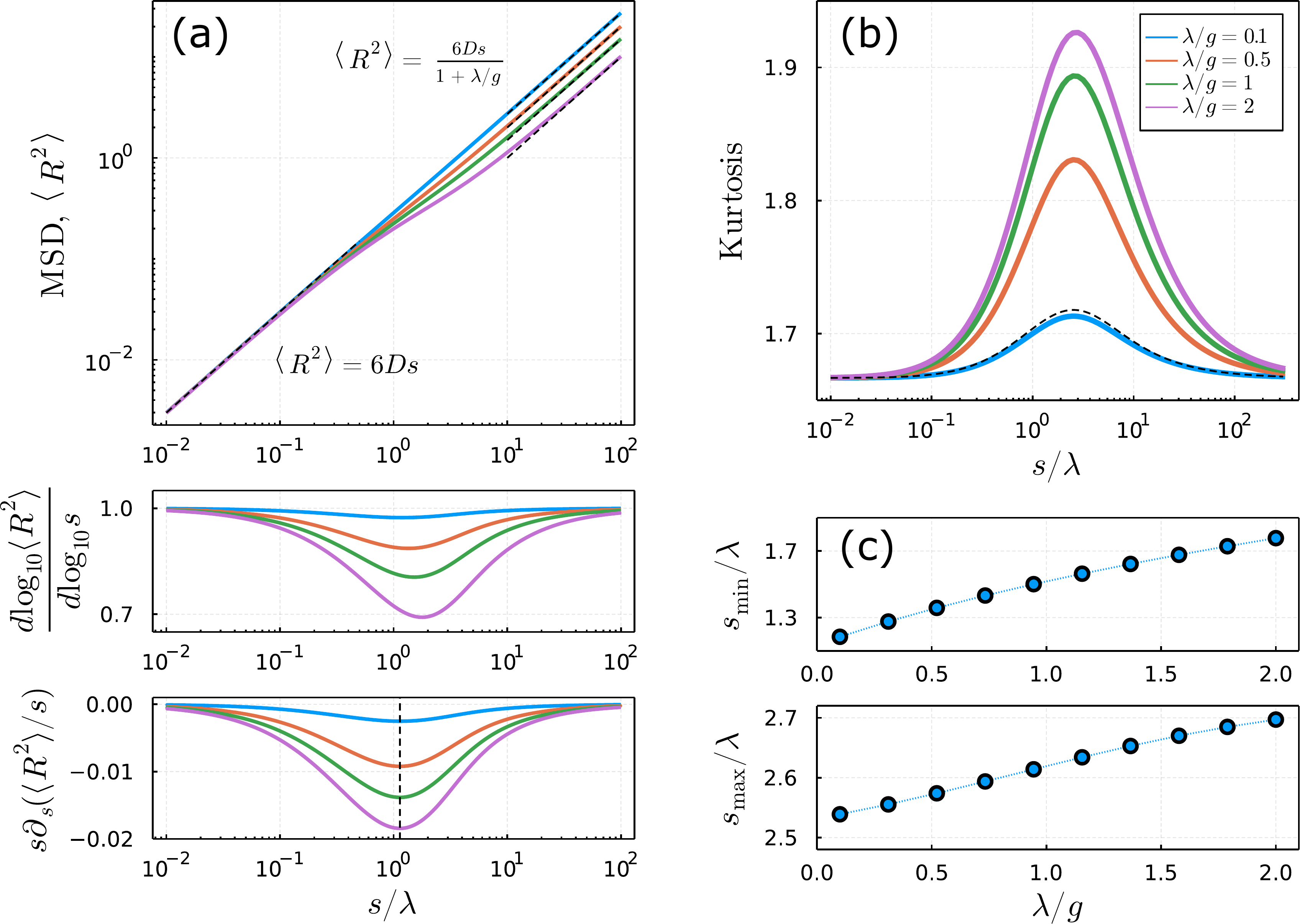}
\caption{
(a) The  MSD $\langle R^2(s) \rangle$ (top panel), its log-log derivative $\frac{d\log_{10}\langle R^2(s) \rangle}{d\log_{10}s}$ (middle panel) and $\frac{d}{ds}\frac{\langle R^2(s)\rangle}{s}$ (bottom panel)  in dependence on the contour separation $s$ for different values of  $\lambda/g$.
 (b) Kurtosis coefficient ${\cal K}(s)$ as a function of contour separation $s$ for the same set of parameters.  (c) The minimum   $s_{\text{min}}$ of the log-log derivative $\frac{d\log_{10}\langle R^2(s) \rangle}{d\log_{10}s}$ (top panel) and the maximum $s_{\text{max}}$ of the kurtosis coefficient  ${\cal K}(s)$ (bottom panel) in their dependence on the dimensionless parameter  $\lambda/g$.}
\label{fig:3}
\end{figure}

Let us pass to the analysis of the asymptotic behavior dictated by Eq. (\ref{msd}).
Since $\lim_{z\to 0} f_{\text{MSD}}(z) = 1$,  we see from Eq. (\ref{msd}) that the well-known ideal-chain scaling law,  $\langle R^2(s)\rangle=l_{\text{eff}}s$, is recovered at $s\ll \lambda$. 
Clearly, this is because the sufficiently small segments of the chain are non-sensitive to the loops constraints.
In the opposite limit one finds $\lim_{z\to \infty} f_{\text{MSD}}(z) = 0$, so that $\langle R^2(s)\rangle=\dfrac{l_{\text{eff}}s}{1+\lambda/g}<l_{\text{eff}}s$ at $s\gg\lambda$ if $\lambda/g\lesssim 1$.
This conclusion  also has rather transparent explanation:  
the random loops compactify the large segments of ideal chain via effective shortening of contour distance between their end points.
As expected, the compactification degree is stronger for larger values of $\lambda/g$.

The double logarithmic scale graph of  $\langle R^2(s)\rangle$ is presented in Fig.~\ref{fig:3}a. 
We see that at $\lambda\lesssim g$  crossover between small- and large-$s$ linear asymptotic regimes takes place at the scale $s\sim \lambda$, whereas the mean inter-loop spacing $g$  affects only the magnitude of disorder-induced perturbation of the MSD profile.
This observation suggests how it would be possible to estimate the average length of the cohesin-anchored loops, having an experimentally measured profile of MSD. 
Namely, analysis of Eq. (\ref{msd}) indicate that the minimum of expression $s\frac{d}{ds}[\frac{\langle R^2(s)\rangle}{s}]$ in its dependence on the contour separation $s$,  is determined by $\lambda$ and is equal to $s_\ast \approx 1.14\lambda$ irrespectively of $g$, see Fig.~\ref{fig:3}a.
Also, it may be informative to analyse the log-derivative  $\frac{d\log_{10}\langle R^2(s) \rangle}{d\log_{10}s}$ which determines the slope of the MSD in the log-log scale plot.
As we see in Fig.~\ref{fig:3}c, the log-derivative exhibits local minimum whose position $s_{min}$ is of the order of $\lambda$ and it changes by only $50\%$ with a twenty-fold increase in $g$.

Beyond the MSD our model allows to  explore how the functional form of the probability density of separation vector $\vec R(s)$ depends on the linear scale  $s$.
Qualitatively, one may expect that  cohesin-mediated random loops do not destroy normality of statistics of sufficiently short chain segments of contour length $s\ll \lambda$ which are not affected by loops constraints.
Also, Gaussianity must also restore at large scales, $s\gg \lambda, g$.
Indeed, for each diagram in Fig. \ref{fig:diagrams} the conditional probability density of $\vec R(s)$ is Gaussian (see Appendix)  with an effective contour separation whose  fluctuations at $s\gg\lambda, g$  become small compared to the average value due to the central limit theorem.


To quantify the  possible deviations  of scale-dependent two-point statistics from Gaussianity we calculate  the kurtosis coefficient defined as  ${\cal K}(s)=\frac{\langle R^4(s)\rangle}{\langle R^2(s)\rangle^2}$.
Clearly, the value $5/3$ corresponds to the normal statistics of three-dimensional ideal chain.
The generalization of non-perturbative calculations presented above to the case of the fourth-order statistical moment $\langle R^4(s)\rangle$, entering the definition of the kurtosis, is possible in principle, but practically difficult to implement.
However, if $\lambda/g\ll 1$ and $s\ll g$, one can neglect the diagrams 
containing two or more cohesin-mediated loops due to
their vanishing statistical weights, and  analytical calculations become feasible.
Expanding statistical weights given by Eqs.~(\ref{eq:Wa})-(\ref{eq:Wc}) in linear order upon small parameters $\lambda/g$ and $s/g$ and using  the relation $ {\cal R}^4_{\alpha}(s| \{ A\}_{\alpha})=\frac53 l_{\text{eff}}^2 \tilde{s}^2_{\alpha}[s,\{ A\}_\alpha]$,  which follows from the Gaussianity of conditional statistics of vector $\vec R$ for each diagram,  we find the following asymptotic result (see Appendix)
\begin{equation}
\label{kurtosis}
{\cal K}(s)=\frac53\frac{\sum_{\alpha}\langle \tilde{s}^2_{\alpha}[s,\{ A\}_\alpha ]\rangle_{loops}}{\sum_{\alpha}\langle \tilde{s}_{\alpha}[s,\{ A\}_\alpha]\rangle_{loops}^2}\approx\frac{5}{3}+\frac{\lambda}{g} f_{\text{Kurt}} \left(\frac{s}{\lambda} \right),
\end{equation}
where $f_{\text{Kurt}}(s)=\frac{2}{3s^2}((9+4s-3s^2)e^{-s}-9+5s+s^2(5+3s){\cal E}_3(s) )$.

Equation (\ref{kurtosis}) tells us that  rare random loops produce a linear correction in small parameter $\lambda/g\ll 1$ to the value $5/3$ corresponding to normal statistics of three-dimensional ideal chain in the absence of loops disorder.
The corresponding plot of the kurtosis coefficient ${\cal K}(s)$ as a function of $s$ is represented in Fig.~\ref{fig:3}b.
Data associated with the regime $\lambda/g\sim1$ were generated via numerical integration of diagram contributions over  exact statistical weights. 
We found that the one-loop prediction (dashed line) is rather accurate at  $\lambda/g\lesssim 0.1$, but underestimates ${\cal K}(s)$ when $\lambda/g\gtrsim 1$. 
In agreement with the general arguments discussed above,  the kurtosis coefficient is close to $5/3$ at $s\ll \lambda$ and $s\gg \lambda$.
At intermediate scales of contour distances statistics of the separation vector $\vec R$ exhibits  deviation from Gaussianity, and
this effect is the more pronounced, the greater the dimensionless parameter $\lambda/g$.
Most importantly, the kurtosis coefficient is peaked at the point $s=s_{max}$, whose position is mainly determined by $\lambda$ and changes by only $10\%$  when $g$ is changed  by a factor of $20$.
Thus, we expect that measurement of the scale-dependent  kurtosis may provide an estimate for mean loop size $\lambda$ along with the analysis of experimental MSD profile.
Note also that the loops-induced violation of normality predicted by our model cannot be reproduced in the framework of Heterogeneous Loop Model \cite{liu2019heterogeneous, liu2021extracting, bak2021unified} 
since it postulates normal statistics of  chromatin  at all genomic scales. The same applies to the modelling approach based on inference of the maximum entropy  distribution of pair-wise distances with experimental mean-squared distances as constraints   \cite{shi2022method}.

\textit{Conclusion.}
To the best to our knowledge, the existing literature lacks the sufficient amount of relevant statistical information  characterised by high genomic and spatial resolution required to directly confront our predictions  with experiment.
Nevertheless, we believe that the required data will become available in the coming years due to modern tools for  high-throughput super-resolution imaging   enabling direct  visualization of the spatial positions of many genomic loci at the single-cell level \cite{hafner2022beyond,gabriele2022dynamics,mach2022live,beckwith2022visualization,sasaki2022quantitative}.
Noteworthy, while modelling the chromatin conformation by ideal chain seem to be reasonable for some types of data  \cite{gassler2017mechanism,polovnikov2022fractal,beckwith2022visualization}, quantitative agreement between theory and experiment in a wider range of situations may require more complex polymer models that resist analytical treatment.
In particular, further (mostly numerical) work is required to establish how  statistics of pairwise distances in the presence of loop extrusion is affected by excluded volume effects.

\acknowledgments

S.B. thanks Leonid A. Mirny,  Hugo B. Brand{\~a}o and Kirill Polovnikov for valuable discussions.
The work was supported by the Russian Science Foundation,
project no. 22-72-10052.

\newpage

\begin{widetext}

\section{Appendix}
The Appendix is structured as follows.  In the first section, we remind the basic statistical properties of free Brownian paths and Brownian bridges  relevant for derivation of diagram contributions. 
Next, in the second section we discuss the basic   properties of two-state Markov jump process required to construct exact statistical  weights of the diagrams. 
In sections III-VI, we derive the integral expressions for the loop-averaged  contributions coming from each type of diagrams. 
Finally, in the section VII, we provide  details of one-loop calculations of the kurtosis coefficient.

\subsection{I. Basic Statistical properties of Brownian paths}

In what follows we will heavily exploit the well-known analogy between a polymer and a random walk, see, e.g., Refs.~\cite{DeGennes_1979,GKh_1994}. Within this analogy, the coordinate along the polymer plays a role of time and the polymer contour is thought of as the trajectory of a random walker, see Fig.~\ref{fig:MCh}a. Adopting this language, we, thus, obtain a random walk whose trajectory represents the alternating free Brownian paths, which correspond to the gap regions of the polymer, and the Brownian bridges corresponding to the cohesion-mediated loops in our original polymer model.
Let us recall the key statistical properties of Brownian motion.



The propagator of the free Brownian motion in three dimensions,
\begin{equation}\label{propagator_BM}
G_{\text{free}}(\vec r,t|\vec r_0,0)=\frac{1}{(4\pi Dt)^{3/2}}\exp\left(-\frac{(\vec r-\vec r_0)^2}{4Dt}\right),
\end{equation}
describes the probability to find the Brownian particle having diffusivity $D$ in the point $\vec r$ after time $t$ if it starts in $\vec r_0$. In context of the polymer model, Eq.~(\ref{propagator_BM}) represents the probability distribution of the separation vector $\vec r-\vec r_0$ between two monomers inside a gap region of the polymer provided that their contour separation is $t$.

The Brownian bridge  is the Brownian trajectory subject to the condition that the particle must return to its starting position after a certain amount of time. Propagator of a Brownian bridge of length $T$ with a base in $\vec r_0$ is given by
\begin{equation}\label{propagator_BB}
G_{\text{bridge}}(\vec r,t|\vec r_0,0;\vec r_0,T)=\frac{G_{\text{free}}(\vec r,t|\vec r_0,0)G_{\text{free}}(\vec r_0,T|\vec r,t)}{G_{\text{free}}(\vec r_0,T|\vec r_0,0)}=\left(\frac{T}{4\pi Dt(T-t)}\right)^{3/2}\exp\left(-\frac{T(\vec r-\vec r_0)^2}{4Dt(T-t)}\right),
\end{equation}
where $0\le t\le T$. Eq.~(\ref{propagator_BB}) describes the probability that the Brownian particle, which starts in $\vec r_0$ and returns to $\vec r_0$ after time $T$, will be in $\vec r$ at the moment of time $t$. Equivalently, this equation defines the probability distribution of the separation vector between the loop base and the monomer inside this loop given the contour separation $t$ and the loop length $T$.

More generally, the Brownian bridge pinned at two different points $\vec r_1$ and $\vec r_2$ at the moments of time $t_1$ and $t_2$, respectively, is characterised by the following probability distribution
\begin{equation}
G_{\text{bridge}}(\vec r,t|\vec r_1,t_1;\vec r_2,t_2)=\left(\frac{t_2-t_1}{4\pi D(t_2-t)(t-t_1)}\right)^{3/2} \exp\left(-\frac{(\vec r_2-\vec r)^2}{4D(t_2-t)}-\frac{(\vec r-\vec r_1)^2}{4D(t-t_1)}+\frac{(\vec r_2-\vec r_1)^2}{4D(t_2-t_1)}\right),
\end{equation}
where $t_1\le t\le t_2$.

In what follows the propagators determined by 
Eqs. (\ref{propagator_BM}) and (\ref{propagator_BB}) play a role of building blocks  of the diagram calculations. 
But before proceeding to the corresponding calculations, we need to discuss the basic properties of the two-state Markov chain  that will be required to derive the statistical weights of the diagrams depicted in Fig. 2 in main text.





\subsection{II. Basic Statistical Properties of Two-State Markov Process}\label{sec:Markov}

\begin{figure}
\includegraphics[width=0.65\linewidth]{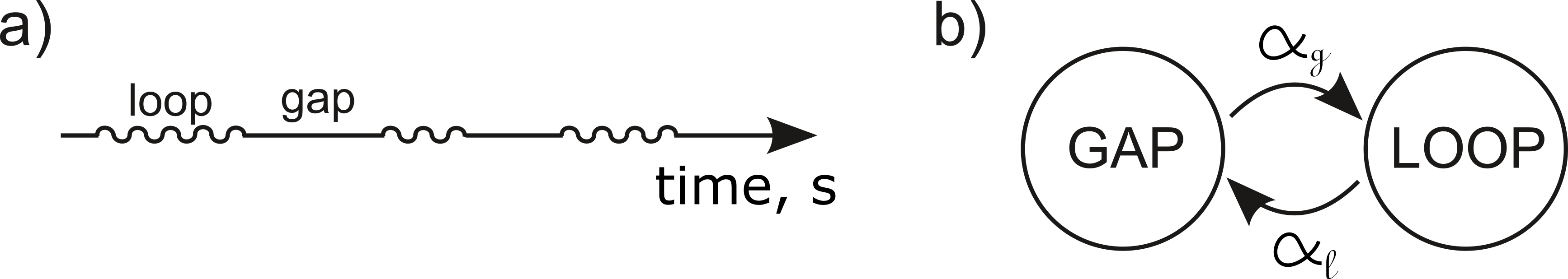}
\caption{ (a) Based on the analogy between polymer conformation and random walk trajectory, we introduce a time axis with time intervals  measured in the units of the polymer contour length. (b) The continuous time Markov jump process with  two states, ``Loop'' and ``Gap'', and transition rates $\alpha_l=\lambda^{-1}$, $\alpha_g=g^{-1}$.
By construction, statistics of time intervals that this auxiliary Markov process spends in different states coincides with the statistics of alternating loops and gaps in our polymer model.}
\label{fig:MCh}
\end{figure} 

Let us consider a Markov process with the transition rates $\alpha_l=1/\lambda$ and $\alpha_g=1/g$ between two states, ``Gap''  and ``Loop'', which dictates the duration of random time intervals which the random walker introduced in previous section spends in the free and looped segments of its trajectory, see Fig.~\ref{fig:MCh}b. In other words, this auxiliary Markov process generates the random length of gaps and loops in the original polymer model. 
 
The stochastic dynamics of the two-state continuous-time Markov jump process  is described by the following pair of equations
\begin{eqnarray}
\label{Markov_F}
\frac{d\pi_g}{ds}&=&-\alpha_g \pi_g+\alpha_l \pi_l,\\
\label{Markov_L}
\frac{d\pi_l}{ds}&=&\alpha_g \pi_g-\alpha_l \pi_l,
\end{eqnarray}
where $\pi_g(s)$ and $\pi_l(s)$ represent the probabilities that the monomer having contour coordinate $s$ lies on the gap or loop, respectively. 
It is straightforward to find the stationary solution of these equations 
\begin{eqnarray}
\label{stationary_sol}
p_g=\frac{\alpha_l}{\alpha_g+\alpha_l}, \qquad 
p_l=\frac{\alpha_g}{\alpha_g+\alpha_l}.
\end{eqnarray}
Clearly, $p_g$ ($p_l$) gives the probability that a randomly chosen point of the polymer with disorder op loops belongs to a gap (loop) region. 

The propagator $\pi_{A\to B}(s)$ of the Markov process is defined as the probability to find the process in the state ``B'' after time $s$ under the condition that it starts in the state ``A''. It is easy to find from Eqs. (\ref{Markov_F}) and (\ref{Markov_L}) that 
\begin{eqnarray}
\label{pi_gg}
&\displaystyle \pi_{g\to g}(s)=\frac{1}{\alpha_g+\alpha_l}\left[\alpha_l+\alpha_g e^{-(\alpha_l+\alpha_g)s}\right],&\\
&\displaystyle \pi_{g\to l}(s)=\frac{\alpha_g}{\alpha_g+\alpha_l}\left[1-e^{-(\alpha_g+\alpha_l)s}\right] ,&\\
&\displaystyle \pi_{l\to g}(s)=\frac{\alpha_l}{\alpha_g+\alpha_l}\left[1-e^{-(\alpha_g+\alpha_l)s}\right],&\\
&\displaystyle \pi_{l\to l}(s)= \frac{1}{\alpha_g+\alpha_l}\left[\alpha_g + \alpha_le^{-(\alpha_g+\alpha_l)s}\right].&
\end{eqnarray}
In the limit $s\to+\infty$, these expressions turn into statistically stationary probabilities $p_g$ and $p_l$ to find the process in given states, i.e. $\lim_{s\to\infty}\pi_{l\to g}(s)=\lim_{s\to\infty}\pi_{g\to g}(s) =p_g$, 
$\lim_{s\to\infty}\pi_{l\to l}(s)=\lim_{s\to\infty}\pi_{g\to l}(s) =p_l$.

To perform averaging over the loop disorder (see below), we will also need to know the statistical moment $\langle x_s\rangle$, where $x_s$ is the time spent in the ``Loop'' state during the time interval $[0, s]$ under the condition that the Markov process occupies the ``Gap'' state at both ends of this interval. 
To calculate the expectation of   $x_s$, we introduce the stochastic variable $\zeta(t)$, which can take two values: $\zeta(t)=l$ if at the moment $t$ the Markov jump process is in the ``Loop'' state, and $\zeta(t)=g$ if the process is currently in the ``Gap'' state. Then the random variable $x_s$ can be represented as
\begin{equation}
x_s= \dfrac{1}{s}\int_0^sI[\zeta(t)=l]dt,
\end{equation}
where $I[...]$ is an indicator variable equal to one if the condition in its argument is true, and equal to zero otherwise. Performing averaging one obtains 
\begin{equation}
\label{x_s_def}
\langle x_s\rangle=\dfrac{1}{s} \int_0^s\langle I[\zeta(t)=l]\rangle dt=\dfrac{1}{s} \int_0^sPr[\zeta(t)=l|\zeta(0)=g,\zeta(s)=g] dt,
\end{equation}
where $Pr[\zeta(t)=l|\zeta(0)=g,\zeta(s)=g]$ is the probability of finding the Markov jump process in the ``Loop'' state at time $t$ given that it was in the ``Gap'' state both at time $0$ and at time $s$. This probability can be easily calculated due to the lack of memory of the past in a Markov process. Indeed, 
\begin{equation}
Pr[\zeta(t)=l|\zeta(0)=g,\zeta(s)=g]=\frac{Pr[\zeta(s)=g|\zeta(t)=l]Pr[\zeta(t)=l|\zeta(0)=g]}{Pr[\zeta(s)=g|\zeta(0)=g]},
\end{equation}
and since $Pr[\zeta(s)=g|\zeta(t)=l]=\pi_{l\to g}(s-t)$, $Pr[\zeta(t)=l|\zeta(0)=g]=\pi_{g\to l}(t)$ and $Pr[\zeta(s)=g|\zeta(0)=g]=\pi_{g\to g}(s)$ we obtain 
\begin{equation}
Pr[\zeta(t)=l|\zeta(0)=g,\zeta(s)=g]=\frac{\pi_{l\to g}(s-t)\pi_{g\to l}(t)}{\pi_{g\to g}(s)}.
\end{equation}
Substituting this result into Eq. (\ref{x_s_def}) yields 
\begin{eqnarray}\label{eq:xs}
\langle x_s\rangle= \dfrac{\alpha_g \alpha_l [2 + (\alpha_g+\alpha_l)s + e^{(\alpha_g + \alpha_l)s} ((\alpha_g+\alpha_l)s-2)]}{s (\alpha_g+\alpha_l)^2 [\alpha_g + \alpha_l e^{(\alpha_g+\alpha_l)s}]}.
\end{eqnarray}

Beyond the mean value, the full  statistics of the random variable $x_s$ can be extracted from the results of Ref.~\cite{Pedler_1971}. Namely, the probability density ${\cal F}(x_s)$ is given by
\begin{equation}
\label{Pedler}
{\cal F}(x_s)=\frac{e^{-\alpha_g s}\delta(x_s)+\sqrt{\frac{\alpha_g\alpha_l(1-x_s)s^2}{x_s}}I_1\left(2\sqrt{\alpha_g\alpha_l x_s(1-x_s)s^2}\right)e^{-\alpha_g(1-x_s)s-\alpha_lx_ss}}{\frac{\alpha_l}{\alpha_g+\alpha_l}+\frac{\alpha_g}{\alpha_g+\alpha_l}e^{-(\alpha_g+\alpha_l)s}},
\end{equation}
where $I_1(...)$ denotes the modified Bessel function of the first kind~\cite{abramowitz1988handbook}. 

\subsection{III. Diagram A. Derivation of Eqs. (3) and (7)}

We wish to calculate  the mean-squared   displacement (MSD) of the random walker after time $s$.
Depending on the modes of the walker motion at the initial and final moments of time we should distinguish four cases represented in Fig.~2 of the main text. 
If the walker is in the free segments of its trajectory both initially and after time $s$, see the diagram in Fig.~2a, then the probability density function of the walker's displacement $\vec r$ is given by the Gaussian distribution
\begin{equation}\label{propagator_GG}
P_{a}(\vec r|s,x_s)=G_{\text{free}}(\vec r,(1-x_s)s|\vec 0,0)= \frac{1}{(4\pi D \tilde s_{a}[s,x_s])^{3/2}}\exp\left(-\frac{r^2}{4D\tilde s_{a}[s,x_s]}\right),
\end{equation}
with the effective contour separation $\tilde s_{a}[s,x_s]=(1-x_s)s$, where  $x_s$  denotes the fraction of time that walker spent performing Brownian bridges during the course of motion; $0\le x_s < 1 $. 
The intuition behind Eq. (\ref{propagator_GG}) is quite transparent: since the closed Brownian paths don't produce the walker's displacement, the overall effect of loops in diagram (a) is equivalent to reduction of the time allowed to the walker for exploration of the neighborhood. For the mean-squared displacement we, thus, obtain
\begin{equation}
{\cal R}^2_{a}(s| x_s)= \int d^3 r r^2 P_{a}(\vec r|s,x_s)=6D \tilde s_{a}[s,x_s].
\end{equation}

Next, using basic properties of two-state Markov jump process described in section II, we find that the diagram $(a)$ is characterized by the following statistical weight 
\begin{equation}\label{eq:Wa}
{\cal W}_{a}(x_s;s)=p_g\pi_{g\to g}(s){\cal F}(x_s),
\end{equation}
where $p_g=\frac{\alpha_l}{\alpha_g+\alpha_l}$ gives the probability that a starting point of the  walker's trajectory belongs to a free Brownian path, $\pi_{g\to g}(s)=\frac{1}{\alpha_g+\alpha_l}(\alpha_l+\alpha_ge^{-(\alpha_l+\alpha_g)s})$ is the probability to find the walker in the free segment of its trajectory after time $s$ under the condition that initially it is also in the free segment, and ${\cal F}(x_s)$ is the probability distribution of the random variable $x_s$ determined by  Eq. (\ref{Pedler}). To average the contribution of the diagram (a) over the disorder of random loops, we should integrate the product ${\cal R}^2_{a}(s| x_s) {\cal W}_{a}(x_s;s)$ over $x_s$ from $0$ up to $1$, i.e.
\begin{equation}
\label{integral_a}
\langle {\cal R}^2_{a}(s| x_s)\rangle_{\text{loops}}=\int_0^1 dx_{s} {\cal R}^2_{a}(s| x_s) {\cal W}_{a}(x_s;s)= 6 D s p_g \pi_{g\to g}(s) (1 - \langle x_s \rangle),
\end{equation}
where $\langle x_s \rangle$ is given by Eq.~(\ref{eq:xs}).

\subsection{IV. Diagram B. Derivation of Eqs. (4) and (8)}

Next, let us assume that the walker starts in the loop and finds itself in the free segment of its trajectory after time $s$. As shown in Fig.~2b of the main text, the loop containing the starting point of the walker's trajectory is parameterized by the time intervals $t_1$ and $t_2$. After averaging over the position of the loop base $\vec r_0$, the probability density function of the walker's displacement becomes
\begin{equation}\label{propagator_LG}
\begin{aligned}
P_{b}(\vec r|s,t_1,t_2,x_{s-t_2})= \int d^3r_0 G_{\text{bridge}}(\vec 0,0|\vec r_0,-t_1;\vec r_0,t_2)  
G_{\text{free}}(\vec r,t_2 + (1-x_{s-t_2})(s-t_2)|\vec r_0,t_2)=\\
=\frac{1}{(4\pi D \tilde{s}_{b}[s,t_1,t_2,x_{s-t_2}])^{3/2}}\exp\left(-\frac{r^2}{4D \tilde{s}_{b}[s,t_1,t_2,x_{s-t_2}]}\right), 
\end{aligned}
\end{equation}
where $\tilde{s}_{b}[s,t_1,t_2,x_{s-t_2}]=(1-x_{s-t_2})(s-t_2)+\frac{t_1t_2}{t_1+t_2}$, and $0\le x_{s-t_2}<1$, $t_1\ge 0$, $0\le t_2 \le s$. Now $x_{s-t_2}$ is the fraction of time the walker spend performing Brownian bridges during the time interval between $t_2$ and $s$. 
Therefore, the mean-squared displacement  of the walker is given by  
\begin{equation}
\label{MSD_diagram_b}
{\cal R}^2_{b}(s|t_1,t_2,x_{s-t_2})= \int d^3 r r^2 P_{b}(\vec r|s,t_1,t_2,x_{s-t_2})=6D \tilde{s}_{b}[s,t_1,t_2,x_{s-t_2}].
\end{equation}

Next, for the statistical weight of the diagram $(b)$ we obtain
\begin{equation}\label{eq:Wb}
{\cal W}_{b}(t_1,t_2,x_{s-t_2};s)=2p_l\alpha^2_l e^{-\alpha_l(t_1+t_2)}\pi_{g\to g}(s-t_2) {\cal F}(x_{s-t_2}),
\end{equation}
where $p_l=\frac{\alpha_g}{\alpha_g+\alpha_l}$ gives the probability that a starting point of the  walker's trajectory belongs to a  loop.
Obviously, the case when the walker starts in the free segment and finishes in the closed segment is completely equivalent to the situation that we have just considered. This explains the origin of factor $2$ in Eq. (\ref{eq:Wb}).

From Eqs. (\ref{MSD_diagram_b}) and (\ref{eq:Wb}) one obtains that the loops-averaged  contribution of the diagram (b) is given by the following   integral  
\begin{equation}\label{integral_b}
\begin{aligned}
  \langle {\cal R}^2_{b}(s| t_1,t_2,x_{s-t_2})\rangle_{\text{loops}}=\int_0^{\infty} dt_1 \int_0^s dt_2  \int_0^1 dx_{s-t_2} {\cal R}^2_{b}(s|t_1,t_2,x_{s-t_2}){\cal W}_{b}(t_1,t_2,x_{s-t_2};s)=\\
  =12 D p_l \alpha_l^2 \int_0^{\infty} dt_1 \int_0^s dt_2 \left[ (1-\langle x_{s-t_2} \rangle)(s-t_2) + \dfrac{t_1 t_2}{t_1 + t_2} \right] e^{-\alpha_l(t_1+t_2)} \pi_{g\to g}(s-t_2). 
\end{aligned}
\end{equation}

\subsection{V. Diagram C. Derivation of Eqs. (5) and (9)}

Now let us consider the scenario when the starting and the final points of the walker's trajectory belong to the same loop, see Fig.~2c in the main text. Performing averaging over the position of the loop base we find the following result for the probability distribution of the walker's displacement after time $s$
\begin{equation}
\begin{aligned}\label{propagator_LL1}
P_{c}(\vec r|s,t_1,t_2)= \int d^3r_0 G_{\text{bridge}}(\vec 0,0|\vec r_0,-t_1;\vec r_0,t_2) G_{\text{bridge}}(\vec r,s|\vec 0,0;\vec r_0,t_2)= \\
=\frac{1}{(4\pi D \tilde{s}_{c}[s,t_1,t_2])^{3/2}}\exp\left(-\frac{r^2}{4D\tilde{s}_{c}[s,t_1,t_2]}\right),
\end{aligned}
\end{equation}
where $\tilde{s}_{c}[s,t_1,t_2]=\left(1-\frac{s}{t_1+t_2}\right)s$, and $t_1\ge 0$, $t_2\ge s$. 
From Eq. (\ref{propagator_LL1}) one obtains  the mean-squared walker's displacement 
\begin{equation}
{\cal R}^2_{c}(s| t_1,t_2)= \int d^3 r r^2 P_{c}(\vec r|s,t_1,t_2)=6D \tilde{s}_{c}[s,t_1,t_2],
\end{equation}
whereas for the the statistical weight of the trajectories described by the diagram $(c)$ we find 
\begin{equation}\label{eq:Wc}
{\cal W}_{c}(t_1,t_2;s)=p_l\alpha_l^2e^{-\alpha_l(t_1+t_2)}.
\end{equation}

Thus, the loop-averaged contribution of the diagram $(c)$ to mean-squared displacement 
is determined by the following double integral
\begin{equation}
\label{integral_c}
\langle {\cal R}^2_{c}(s| t_1,t_2)\rangle_{\text{loops}}=\int\limits_0^{\infty} dt_1 \int\limits_s^{\infty} dt_2 {\cal R}^2_{c}(s| t_1,t_2){\cal W}_{c}(t_1,t_2;s)= 6Ds p_l\alpha_l^2 \int\limits_0^{\infty} dt_1 \int\limits_s^{\infty} dt_2 \left( 1 - \dfrac{s}{t_1+t_2}\right) e^{-\alpha_l(t_1+t_2)}.
\end{equation}

\subsection{VI. Diagram D. Derivation of Eqs. (6) and (10)}

Finally, the probability distribution of the walker's displacement in the situation when the initial and final point of its trajectory belong to different loops is given by
\begin{equation}
\label{propagator_LL2}
P_{d}(\vec r|s,t_1,t_2,\tau,x_\tau,T_2)=\frac{1}{(4\pi D \tilde{s}_{d}[s,t_1,t_2,\tau,x_{\tau},\tilde{T}]}\exp\left(-\frac{r^2}{4D\tilde{s}_{d}[s,t_1,t_2,\tau,x_{\tau},\tilde{T}]}\right),
\end{equation}
where $\tilde{s}_{d}[s,t_1,t_2,\tau,x_{\tau},\tilde{T}]=\frac{t_1t_2}{t_1+t_2}+(1-x_\tau)\tau+\frac{\tilde{t}_1\tilde{t}_2}{\tilde{t}_1+\tilde{t}_2}$,
and $\tilde{t}_1=\tilde{T}+\tau+t_2-s$, $\tilde{t}_2=s-\tau-t_2$, $t_1\ge 0$, $0\le t_2\le s$, $0\le\tau\le s-t_2$, $0\le x_\tau\le 1$, $\tilde{T}\ge s-t_2-\tau$. In this case, $x_\tau$ denotes the fraction of time the walker spend in "Loop"\  state during the time interval between $t_2$ and $s-\tilde{t}_2$. From Eq. (\ref{propagator_LL2}) one obtains  the mean-squared walker's displacement 
\begin{equation}
{\cal R}^2_{d}(s| t_1,t_2,\tau,x_\tau,\tilde T)= \int d^3 r r^2 P_{c}(\vec r|s,t_1,t_2)=6D \tilde{s}_{d}[s,t_1,t_2,\tau,x_{\tau},\tilde{T}].
\end{equation}

Clearly, the statistical weight of the trajectories described by the diagram $(d)$ is given by
\begin{eqnarray}
\label{eq:Wd}
{\cal W}_{d}(t_1,t_2,\tau,x_\tau,\tilde T;s)=p_l\alpha_l^3e^{-\alpha_l(t_1+t_2+\tilde T)}\alpha_g \pi_{g\to g}(\tau) {\cal F}(x_\tau).
\end{eqnarray}

Thus, for the loops-averaged contribution of the diagram (d) we find  
\begin{equation}\label{integral_d}
\begin{aligned}
    \langle {\cal R}^2_{d}(s|  t_1,t_2,\tau,x_\tau,\tilde T)\rangle_{\text{loops}}=\int\limits_0^{\infty} dt_1 \int\limits_0^s dt_2 \int\limits_0^{s-t_2} d \tau \hspace*{-0.2cm} \int\limits_{s-t_2-\tau}^{\infty} \hspace*{-0.2cm} d \tilde T\int\limits_0^1dx_\tau {\cal R}^2_{d}(s| t_1,t_2,\tau,x_\tau,\tilde T){\cal W}_{d}(t_1,t_2,\tau,x_\tau,\tilde T;s) =\\
6D p_l\alpha_l^3 \alpha_g \int\limits_0^{\infty} dt_1 \int\limits_0^s dt_2 \hspace*{-0.1cm} \int\limits_0^{s-t_2} \hspace*{-0.1cm} d \tau \hspace*{-0.3cm} \int\limits_{s-t_2-\tau}^{\infty} \hspace*{-0.3cm} d \tilde T \left[ (1-\langle x_{\tau} \rangle) \tau + \dfrac{t_1 t_2}{t_1 + t_2} + \dfrac{(\tilde T +t_2+\tau-s)(s-t_2-\tau)}{\tilde T} \right] e^{-\alpha_l(t_1+t_2+\tilde T)} \pi_{g\to g}(\tau).
\end{aligned}
\end{equation}

Calculating the integrals in Eqs. (\ref{integral_a}), (\ref{integral_b}), (\ref{integral_c}) and (\ref{integral_d})  and summing the resulting expressions, we arrive at the Eq. (12) in main text.

\subsection{VII. One-loop approximation. Derivation of Eq. (13)}

The kurtosis coefficient of the random vector $\vec R(s)$ is defined as 
\begin{equation}
\label{kurtosis_definition}
{\cal K}(s)=\frac{\langle R^4(s)\rangle}{\langle R^2(s)\rangle^2}.
\end{equation}
We already know that the MSD $\langle R^2(s)\rangle$ in our model is given by Eq. (12) in main text.
As for the fourth order statistical moment $\langle R^4(s)\rangle$, 
taking into account the Gaussian form of the conditional distribution functions (\ref{propagator_GG}), (\ref{propagator_LG}), (\ref{propagator_LL1}), and (\ref{propagator_LL2}), one readily obtains
\begin{equation}
\label{R4_d_exp}
\langle R^4(s)\rangle=60D^2\sum_{\alpha=a,b,c,d}\langle \tilde{s}^2_{\alpha}[s,\{ A\}_\alpha]\rangle_{\text{loops}}.
\end{equation}




Exact diagrammatic calculations accordingly to Eq. (\ref{R4_d_exp}) are possible in principle, but practically difficult to implement.
However, analytical derivation of the fourth moment $\langle R^4(s)\rangle$ becomes feasible in the rare loops limit.
More specifically, 
if  $\lambda/g\ll 1$ and $s/g\ll 1$, then  one can neglect the realizations of diagrams where there is more than one loop, see Fig.~\ref{fig:one-loop}.
Then, neglecting the diagram (d) and simplifying the formulas (3)-(5) from the main text, we obtain
\begin{equation}
\langle R^4(s)\rangle\approx 60D^2\sum_{\alpha=a,b,c}\langle \tilde{s}^2_{\alpha}[s,\{ A\}_\alpha]\rangle_{\text{one loop}},
\end{equation}
where 
\begin{eqnarray}
\label{s_eff_a_one}
&\tilde{s}_{a}[s,x_s]=(1-x_s)s,&\\
\label{s_eff_b_one}
&\tilde{s}_{b}[s,t_1,t_2]=s-t_2+\frac{t_1t_2}{t_1+t_2},&\\
\label{s_eff_c_one}
&\tilde{s}_{c}[s,t_1,t_2]=\left(1-\frac{s}{t_1+t_2}\right)s.&
\end{eqnarray}


\begin{figure}
\includegraphics[width=0.75\linewidth]{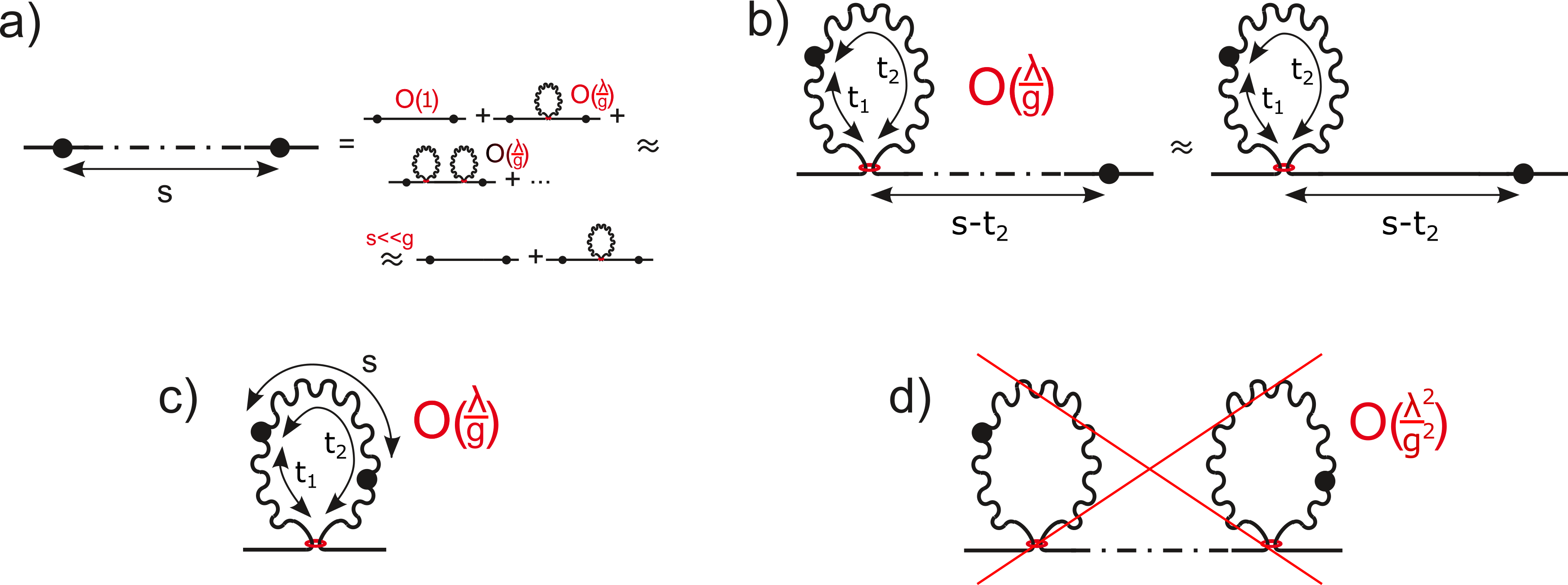}
\caption{For $\lambda/g\ll 1$ and $s/g\ll 1$ the two-point statistics of an ideal chain with a disorder of random loops can be computed in the one-loop approximation, leaving only those diagrams containing at most one cohesin-mediated loop. In other words, diagram (d) can simply be ignored, and the dash-dotted line in diagrams (a) and (b) can be replaced by a solid line.}
\label{fig:one-loop}
\end{figure} 

When averaging over the disorder of the loops, it is convenient to pass to the new variables $T$ and $q$ defined as 
\begin{equation}
t_1=(1-q)T,\qquad t_2=qT.
\end{equation} 
In terms of these variables,  the  diagrams (a), (b) and (c) depicted in Fig. \ref{fig:one-loop} are characterized by the following statistical weights 
\begin{eqnarray}
\label{W_a_simp}
&{\cal W}_{a}(x_s|s)=p_g \pi_{g\to g}(s){\cal F}(x_s), \ \ \text{for} \ \ 0\le x_s< 1,&\\
\label{W_b_simp}
&{\cal W}_{b}(T,q,x_{s-qT}|s)=2p_l\tilde \rho_l(T) \pi_{g\to g}(s-qT) {\cal F}(x_{s-qT}), \ \ \text{for} \ \ 0\le q\le min[1,\frac{s}{T}], T\ge 0, 0\le x_{s-qT}< 1,&\\
\label{W_c_simp}
&{\cal W}_{c}(T,q|s)=p_l\tilde \rho_l(T), \ \ \text{for} \ \  \frac{s}{T}\le q\le 1, T\ge s,&
\end{eqnarray}
where 
$\tilde \rho_l(T)$ denote the probability density of the random loop length in the statistical experiment where loops are sampled by random choice of points along the polymer.
Clearly, $\tilde \rho_l(T)=\frac{T}{\lambda}\rho_l(T)$, where $\rho_l(T)=\frac{1}{\lambda}\exp(-\frac{T}{\lambda})$ is the actual loop length distribution.

Using the smallness of the dimensionless parameters $\lambda/g\ll 1$ and $s/g\ll 1$, we find from Eqs. (\ref{stationary_sol}), (\ref{pi_gg}) and (\ref{Pedler})
\begin{equation}
\label{p_St}
p_g=\frac{g}{g+\lambda}\approx 1-\frac{\lambda}{g}, \qquad p_l=\frac{\lambda}{g+\lambda}\approx\frac{\lambda}{g},
\end{equation}  
and 
\begin{equation}
\label{F(x|s)}
\pi_{g\to g}(s){\cal F}(x_s)\approx(1-\frac{s}{g})\delta(x_s)+\frac{(1-x_s)s^2}{g}\rho_l(x_ss).
\end{equation}
By inserting Eqs. (\ref{p_St}) and (\ref{F(x|s)}) into Eqs. (\ref{W_a_simp},\ref{W_b_simp},\ref{W_c_simp}) and neglecting the  terms nonlinear in the parameter $\lambda/g$ one obtains 
 \begin{eqnarray}
 \label{Wa}
&{\cal W}_{a}(x_s|s)\approx\delta(x_s)+\frac{\lambda}{g}\left(-(1+\frac{s}{\lambda})\delta(x_s)+\frac{(1-x_s)s^2}{\lambda}\rho_l(x_ss)\right), \ \ \text{for} \ \ 0\le x_s< 1,& \\
 \label{Wb}
 &{\cal W}_{b}(T,q|s)\approx2\frac{\lambda}{g}\frac{T}{\lambda}\rho_l(T), \ \ \text{for} \ \ 0\le q\le min[1,\frac{s}{T}], T\ge 0,&\\
  \label{Wc}
&{\cal W}_{c}(T,q|s)\approx\frac{\lambda}{g}\frac{T}{\lambda}\rho_l(T),  \ \ \text{for} \ \  \frac{s}{T}\le q\le 1, T\ge s.&
 \end{eqnarray}


Next, performing averaging over disorder of loops, we find that in the first order-approximation with respect to the ratio $\lambda/g$,  the fourth-order statistical moment of the random vector $\Vec R(s)$ is given by
\begin{eqnarray}
\label{R4_one-loop}
    &&\langle R^4(s)\rangle\approx 60D^2\int_0^1dx_s \tilde s^2_a[s,x_s]{\cal W}_a(x_s|s)+60D^2\int_0^\infty dT \int_0^{min[1,s/T]} dq \tilde s^2_b[s,T,q]{\cal W}_b(T,q|s) + \\
    &&+60D^2\int_s^\infty dT \int_{s/T}^{1} dq \tilde s^2_c[s,T,q]{\cal W}_c(T,q|s)=\\
    &&=60(Ds)^2\left(1+\frac{\lambda}{g}\frac{s^2}{\lambda}\left[\int_{0}^1dx \rho_l(xs)(-\frac{3}{5}x^3+\frac{5}{3}x^2-2x)+ \int_{1}^{+\infty}dx \rho_l(xs)(-\frac{3}{5x^2}+\frac{5}{3x}-2)\right] \right)=\\
    &&=60(Ds)^2\left(1+\frac{\lambda}{g}f_4(\frac{s}{\lambda}) \right),
\end{eqnarray}
where 
\begin{equation}
f_4(s)=\frac{-54-96e^{-s}-10s(3s-5)+24(25+9s){\cal E}_5(s)}{15s^2},
\end{equation}
and ${\cal E}_n(s)=\int_{1}^{+\infty}x^{-n}e^{-sx}dx$ is the exponential integral function.

As follows from  Eq. (12) in the main text, the MSD in the same approximation is given by 
\begin{eqnarray}
\label{R2_one-loop}
    \langle R^2(s)\rangle\approx 6Ds\left[1+\frac{\lambda}{d}\left(\frac{2\lambda(1-e^{-\frac{s}{\lambda}})}{3s}-1+\frac{2}{3}{\cal E}_3(\frac{s}{\lambda})\right)\right]
\end{eqnarray}

Substituting Eqs. (\ref{R4_one-loop}) and (\ref{R2_one-loop}) into Eq. (\ref{kurtosis_definition}) finally yields 
\begin{equation}\label{eq:Kurtosis_SM}
{\cal K}(s)\approx\frac{5}{3}+\frac{\lambda}{g} f_{\text{Kurt}}(\frac{s}{\lambda}),
\end{equation}
where
\begin{equation}
f_{\text{Kurt}}(s)=\frac{2}{3s^2}\left((9+4s-3s^2)e^{-s}-9+5s+s^2(5+3s){\cal E}_3(s) \right).
\end{equation}
This result matches equation (13) from the main text.



As noted above, the one-loop approximation relies on smallness of two dimensionless parameters: $\lambda/g$ and $s/g$.
However, by a happy coincidence  the one-loop answer for MSD agrees with the exact result given by Eq.~(12) in the main text for arbitrary large value of $s/g$ provided $\lambda/g\ll 1$. In other words, the large-scale behaviour of MSD obtained from one-loop calculations is accurate for any value of $s$, despite the one-loop approximation is justified only if $s \ll g$. This fact allows us to conclude that since the statistics of zero-mean random vector $\vec R(s)$ is Gaussian at $s \gg g, \lambda$, the one-loop prediction for kurtosis given by Eq. (\ref{eq:Kurtosis_SM}) also remains valid for arbitrary $s$ when $\lambda/g \ll 1$. 

\end{widetext}

\bibliographystyle{apsrev4-2}
\bibliography{Arxiv.bib}

\begin{thebibliography}{58}%
\makeatletter
\providecommand \@ifxundefined [1]{%
 \@ifx{#1\undefined}
}%
\providecommand \@ifnum [1]{%
 \ifnum #1\expandafter \@firstoftwo
 \else \expandafter \@secondoftwo
 \fi
}%
\providecommand \@ifx [1]{%
 \ifx #1\expandafter \@firstoftwo
 \else \expandafter \@secondoftwo
 \fi
}%
\providecommand \natexlab [1]{#1}%
\providecommand \enquote  [1]{``#1''}%
\providecommand \bibnamefont  [1]{#1}%
\providecommand \bibfnamefont [1]{#1}%
\providecommand \citenamefont [1]{#1}%
\providecommand \href@noop [0]{\@secondoftwo}%
\providecommand \href [0]{\begingroup \@sanitize@url \@href}%
\providecommand \@href[1]{\@@startlink{#1}\@@href}%
\providecommand \@@href[1]{\endgroup#1\@@endlink}%
\providecommand \@sanitize@url [0]{\catcode `\\12\catcode `\$12\catcode
  `\&12\catcode `\#12\catcode `\^12\catcode `\_12\catcode `\%12\relax}%
\providecommand \@@startlink[1]{}%
\providecommand \@@endlink[0]{}%
\providecommand \url  [0]{\begingroup\@sanitize@url \@url }%
\providecommand \@url [1]{\endgroup\@href {#1}{\urlprefix }}%
\providecommand \urlprefix  [0]{URL }%
\providecommand \Eprint [0]{\href }%
\providecommand \doibase [0]{https://doi.org/}%
\providecommand \selectlanguage [0]{\@gobble}%
\providecommand \bibinfo  [0]{\@secondoftwo}%
\providecommand \bibfield  [0]{\@secondoftwo}%
\providecommand \translation [1]{[#1]}%
\providecommand \BibitemOpen [0]{}%
\providecommand \bibitemStop [0]{}%
\providecommand \bibitemNoStop [0]{.\EOS\space}%
\providecommand \EOS [0]{\spacefactor3000\relax}%
\providecommand \BibitemShut  [1]{\csname bibitem#1\endcsname}%
\let\auto@bib@innerbib\@empty
\bibitem [{\citenamefont {Ganji}\ \emph {et~al.}(2018)\citenamefont {Ganji},
  \citenamefont {Shaltiel}, \citenamefont {Bisht}, \citenamefont {Kim},
  \citenamefont {Kalichava}, \citenamefont {Haering},\ and\ \citenamefont
  {Dekker}}]{ganji2018real}%
  \BibitemOpen
  \bibfield  {author} {\bibinfo {author} {\bibfnamefont {M.}~\bibnamefont
  {Ganji}}, \bibinfo {author} {\bibfnamefont {I.~A.}\ \bibnamefont {Shaltiel}},
  \bibinfo {author} {\bibfnamefont {S.}~\bibnamefont {Bisht}}, \bibinfo
  {author} {\bibfnamefont {E.}~\bibnamefont {Kim}}, \bibinfo {author}
  {\bibfnamefont {A.}~\bibnamefont {Kalichava}}, \bibinfo {author}
  {\bibfnamefont {C.~H.}\ \bibnamefont {Haering}},\ and\ \bibinfo {author}
  {\bibfnamefont {C.}~\bibnamefont {Dekker}},\ }\href@noop {} {\bibfield
  {journal} {\bibinfo  {journal} {Science}\ }\textbf {\bibinfo {volume}
  {360}},\ \bibinfo {pages} {102} (\bibinfo {year} {2018})}\BibitemShut
  {NoStop}%
\bibitem [{\citenamefont {Golfier}\ \emph {et~al.}(2020)\citenamefont
  {Golfier}, \citenamefont {Quail}, \citenamefont {Kimura},\ and\ \citenamefont
  {Brugu{\'e}s}}]{golfier2020cohesin}%
  \BibitemOpen
  \bibfield  {author} {\bibinfo {author} {\bibfnamefont {S.}~\bibnamefont
  {Golfier}}, \bibinfo {author} {\bibfnamefont {T.}~\bibnamefont {Quail}},
  \bibinfo {author} {\bibfnamefont {H.}~\bibnamefont {Kimura}},\ and\ \bibinfo
  {author} {\bibfnamefont {J.}~\bibnamefont {Brugu{\'e}s}},\ }\href@noop {}
  {\bibfield  {journal} {\bibinfo  {journal} {Elife}\ }\textbf {\bibinfo
  {volume} {9}},\ \bibinfo {pages} {e53885} (\bibinfo {year}
  {2020})}\BibitemShut {NoStop}%
\bibitem [{\citenamefont {Kong}\ \emph {et~al.}(2020)\citenamefont {Kong},
  \citenamefont {Cutts}, \citenamefont {Pan}, \citenamefont {Beuron},
  \citenamefont {Kaliyappan}, \citenamefont {Xue}, \citenamefont {Morris},
  \citenamefont {Musacchio}, \citenamefont {Vannini},\ and\ \citenamefont
  {Greene}}]{kong2020human}%
  \BibitemOpen
  \bibfield  {author} {\bibinfo {author} {\bibfnamefont {M.}~\bibnamefont
  {Kong}}, \bibinfo {author} {\bibfnamefont {E.~E.}\ \bibnamefont {Cutts}},
  \bibinfo {author} {\bibfnamefont {D.}~\bibnamefont {Pan}}, \bibinfo {author}
  {\bibfnamefont {F.}~\bibnamefont {Beuron}}, \bibinfo {author} {\bibfnamefont
  {T.}~\bibnamefont {Kaliyappan}}, \bibinfo {author} {\bibfnamefont
  {C.}~\bibnamefont {Xue}}, \bibinfo {author} {\bibfnamefont {E.~P.}\
  \bibnamefont {Morris}}, \bibinfo {author} {\bibfnamefont {A.}~\bibnamefont
  {Musacchio}}, \bibinfo {author} {\bibfnamefont {A.}~\bibnamefont {Vannini}},\
  and\ \bibinfo {author} {\bibfnamefont {E.~C.}\ \bibnamefont {Greene}},\
  }\href@noop {} {\bibfield  {journal} {\bibinfo  {journal} {Molecular cell}\
  }\textbf {\bibinfo {volume} {79}},\ \bibinfo {pages} {99} (\bibinfo {year}
  {2020})}\BibitemShut {NoStop}%
\bibitem [{\citenamefont {Davidson}\ \emph {et~al.}(2019)\citenamefont
  {Davidson}, \citenamefont {Bauer}, \citenamefont {Goetz}, \citenamefont
  {Tang}, \citenamefont {Wutz},\ and\ \citenamefont
  {Peters}}]{davidson2019dna}%
  \BibitemOpen
  \bibfield  {author} {\bibinfo {author} {\bibfnamefont {I.~F.}\ \bibnamefont
  {Davidson}}, \bibinfo {author} {\bibfnamefont {B.}~\bibnamefont {Bauer}},
  \bibinfo {author} {\bibfnamefont {D.}~\bibnamefont {Goetz}}, \bibinfo
  {author} {\bibfnamefont {W.}~\bibnamefont {Tang}}, \bibinfo {author}
  {\bibfnamefont {G.}~\bibnamefont {Wutz}},\ and\ \bibinfo {author}
  {\bibfnamefont {J.-M.}\ \bibnamefont {Peters}},\ }\href@noop {} {\bibfield
  {journal} {\bibinfo  {journal} {Science}\ }\textbf {\bibinfo {volume}
  {366}},\ \bibinfo {pages} {1338} (\bibinfo {year} {2019})}\BibitemShut
  {NoStop}%
\bibitem [{\citenamefont {Kim}\ \emph {et~al.}(2019)\citenamefont {Kim},
  \citenamefont {Shi}, \citenamefont {Zhang}, \citenamefont {Finkelstein},\
  and\ \citenamefont {Yu}}]{kim2019human}%
  \BibitemOpen
  \bibfield  {author} {\bibinfo {author} {\bibfnamefont {Y.}~\bibnamefont
  {Kim}}, \bibinfo {author} {\bibfnamefont {Z.}~\bibnamefont {Shi}}, \bibinfo
  {author} {\bibfnamefont {H.}~\bibnamefont {Zhang}}, \bibinfo {author}
  {\bibfnamefont {I.~J.}\ \bibnamefont {Finkelstein}},\ and\ \bibinfo {author}
  {\bibfnamefont {H.}~\bibnamefont {Yu}},\ }\href@noop {} {\bibfield  {journal}
  {\bibinfo  {journal} {Science}\ }\textbf {\bibinfo {volume} {366}},\ \bibinfo
  {pages} {1345} (\bibinfo {year} {2019})}\BibitemShut {NoStop}%
\bibitem [{\citenamefont {Ryu}\ \emph {et~al.}(2020)\citenamefont {Ryu},
  \citenamefont {Katan}, \citenamefont {van~der Sluis}, \citenamefont {Wisse},
  \citenamefont {de~Groot}, \citenamefont {Haering},\ and\ \citenamefont
  {Dekker}}]{ryu2020condensin}%
  \BibitemOpen
  \bibfield  {author} {\bibinfo {author} {\bibfnamefont {J.-K.}\ \bibnamefont
  {Ryu}}, \bibinfo {author} {\bibfnamefont {A.~J.}\ \bibnamefont {Katan}},
  \bibinfo {author} {\bibfnamefont {E.~O.}\ \bibnamefont {van~der Sluis}},
  \bibinfo {author} {\bibfnamefont {T.}~\bibnamefont {Wisse}}, \bibinfo
  {author} {\bibfnamefont {R.}~\bibnamefont {de~Groot}}, \bibinfo {author}
  {\bibfnamefont {C.~H.}\ \bibnamefont {Haering}},\ and\ \bibinfo {author}
  {\bibfnamefont {C.}~\bibnamefont {Dekker}},\ }\href@noop {} {\bibfield
  {journal} {\bibinfo  {journal} {Nature Structural \& Molecular Biology}\
  }\textbf {\bibinfo {volume} {27}},\ \bibinfo {pages} {1134} (\bibinfo {year}
  {2020})}\BibitemShut {NoStop}%
\bibitem [{\citenamefont {Banigan}\ and\ \citenamefont
  {Mirny}(2020)}]{banigan2020loop}%
  \BibitemOpen
  \bibfield  {author} {\bibinfo {author} {\bibfnamefont {E.~J.}\ \bibnamefont
  {Banigan}}\ and\ \bibinfo {author} {\bibfnamefont {L.~A.}\ \bibnamefont
  {Mirny}},\ }\href@noop {} {\bibfield  {journal} {\bibinfo  {journal} {Current
  opinion in cell biology}\ }\textbf {\bibinfo {volume} {64}},\ \bibinfo
  {pages} {124} (\bibinfo {year} {2020})}\BibitemShut {NoStop}%
\bibitem [{\citenamefont {Kimura}\ \emph {et~al.}(1999)\citenamefont {Kimura},
  \citenamefont {Rybenkov}, \citenamefont {Crisona}, \citenamefont {Hirano},\
  and\ \citenamefont {Cozzarelli}}]{kimura199913s}%
  \BibitemOpen
  \bibfield  {author} {\bibinfo {author} {\bibfnamefont {K.}~\bibnamefont
  {Kimura}}, \bibinfo {author} {\bibfnamefont {V.~V.}\ \bibnamefont
  {Rybenkov}}, \bibinfo {author} {\bibfnamefont {N.~J.}\ \bibnamefont
  {Crisona}}, \bibinfo {author} {\bibfnamefont {T.}~\bibnamefont {Hirano}},\
  and\ \bibinfo {author} {\bibfnamefont {N.~R.}\ \bibnamefont {Cozzarelli}},\
  }\href@noop {} {\bibfield  {journal} {\bibinfo  {journal} {Cell}\ }\textbf
  {\bibinfo {volume} {98}},\ \bibinfo {pages} {239} (\bibinfo {year}
  {1999})}\BibitemShut {NoStop}%
\bibitem [{\citenamefont {Nasmyth}(2001)}]{nasmyth2001disseminating}%
  \BibitemOpen
  \bibfield  {author} {\bibinfo {author} {\bibfnamefont {K.}~\bibnamefont
  {Nasmyth}},\ }\href@noop {} {\bibfield  {journal} {\bibinfo  {journal}
  {Annual review of genetics}\ }\textbf {\bibinfo {volume} {35}},\ \bibinfo
  {pages} {673} (\bibinfo {year} {2001})}\BibitemShut {NoStop}%
\bibitem [{\citenamefont {Riggs}(1990)}]{riggs1990dna}%
  \BibitemOpen
  \bibfield  {author} {\bibinfo {author} {\bibfnamefont {A.}~\bibnamefont
  {Riggs}},\ }\href@noop {} {\bibfield  {journal} {\bibinfo  {journal}
  {Philosophical Transactions of the Royal Society of London. B, Biological
  Sciences}\ }\textbf {\bibinfo {volume} {326}},\ \bibinfo {pages} {285}
  (\bibinfo {year} {1990})}\BibitemShut {NoStop}%
\bibitem [{\citenamefont {Sanborn}\ \emph {et~al.}(2015)\citenamefont
  {Sanborn}, \citenamefont {Rao}, \citenamefont {Huang}, \citenamefont
  {Durand}, \citenamefont {Huntley}, \citenamefont {Jewett}, \citenamefont
  {Bochkov}, \citenamefont {Chinnappan}, \citenamefont {Cutkosky},
  \citenamefont {Li} \emph {et~al.}}]{sanborn2015chromatin}%
  \BibitemOpen
  \bibfield  {author} {\bibinfo {author} {\bibfnamefont {A.~L.}\ \bibnamefont
  {Sanborn}}, \bibinfo {author} {\bibfnamefont {S.~S.}\ \bibnamefont {Rao}},
  \bibinfo {author} {\bibfnamefont {S.-C.}\ \bibnamefont {Huang}}, \bibinfo
  {author} {\bibfnamefont {N.~C.}\ \bibnamefont {Durand}}, \bibinfo {author}
  {\bibfnamefont {M.~H.}\ \bibnamefont {Huntley}}, \bibinfo {author}
  {\bibfnamefont {A.~I.}\ \bibnamefont {Jewett}}, \bibinfo {author}
  {\bibfnamefont {I.~D.}\ \bibnamefont {Bochkov}}, \bibinfo {author}
  {\bibfnamefont {D.}~\bibnamefont {Chinnappan}}, \bibinfo {author}
  {\bibfnamefont {A.}~\bibnamefont {Cutkosky}}, \bibinfo {author}
  {\bibfnamefont {J.}~\bibnamefont {Li}}, \emph {et~al.},\ }\href@noop {}
  {\bibfield  {journal} {\bibinfo  {journal} {Proceedings of the National
  Academy of Sciences}\ }\textbf {\bibinfo {volume} {112}},\ \bibinfo {pages}
  {E6456} (\bibinfo {year} {2015})}\BibitemShut {NoStop}%
\bibitem [{\citenamefont {Fudenberg}\ \emph {et~al.}(2016)\citenamefont
  {Fudenberg}, \citenamefont {Imakaev}, \citenamefont {Lu}, \citenamefont
  {Goloborodko}, \citenamefont {Abdennur},\ and\ \citenamefont
  {Mirny}}]{fudenberg2016formation}%
  \BibitemOpen
  \bibfield  {author} {\bibinfo {author} {\bibfnamefont {G.}~\bibnamefont
  {Fudenberg}}, \bibinfo {author} {\bibfnamefont {M.}~\bibnamefont {Imakaev}},
  \bibinfo {author} {\bibfnamefont {C.}~\bibnamefont {Lu}}, \bibinfo {author}
  {\bibfnamefont {A.}~\bibnamefont {Goloborodko}}, \bibinfo {author}
  {\bibfnamefont {N.}~\bibnamefont {Abdennur}},\ and\ \bibinfo {author}
  {\bibfnamefont {L.~A.}\ \bibnamefont {Mirny}},\ }\href@noop {} {\bibfield
  {journal} {\bibinfo  {journal} {Cell reports}\ }\textbf {\bibinfo {volume}
  {15}},\ \bibinfo {pages} {2038} (\bibinfo {year} {2016})}\BibitemShut
  {NoStop}%
\bibitem [{\citenamefont {Fudenberg}\ \emph {et~al.}(2017)\citenamefont
  {Fudenberg}, \citenamefont {Abdennur}, \citenamefont {Imakaev}, \citenamefont
  {Goloborodko},\ and\ \citenamefont {Mirny}}]{fudenberg2017emerging}%
  \BibitemOpen
  \bibfield  {author} {\bibinfo {author} {\bibfnamefont {G.}~\bibnamefont
  {Fudenberg}}, \bibinfo {author} {\bibfnamefont {N.}~\bibnamefont {Abdennur}},
  \bibinfo {author} {\bibfnamefont {M.}~\bibnamefont {Imakaev}}, \bibinfo
  {author} {\bibfnamefont {A.}~\bibnamefont {Goloborodko}},\ and\ \bibinfo
  {author} {\bibfnamefont {L.~A.}\ \bibnamefont {Mirny}},\ }in\ \href@noop {}
  {\emph {\bibinfo {booktitle} {Cold Spring Harbor symposia on quantitative
  biology}}},\ Vol.~\bibinfo {volume} {82}\ (\bibinfo {organization} {Cold
  Spring Harbor Laboratory Press},\ \bibinfo {year} {2017})\ pp.\ \bibinfo
  {pages} {45--55}\BibitemShut {NoStop}%
\bibitem [{\citenamefont {Mirny}\ \emph {et~al.}(2019)\citenamefont {Mirny},
  \citenamefont {Imakaev},\ and\ \citenamefont {Abdennur}}]{mirny2019two}%
  \BibitemOpen
  \bibfield  {author} {\bibinfo {author} {\bibfnamefont {L.~A.}\ \bibnamefont
  {Mirny}}, \bibinfo {author} {\bibfnamefont {M.}~\bibnamefont {Imakaev}},\
  and\ \bibinfo {author} {\bibfnamefont {N.}~\bibnamefont {Abdennur}},\
  }\href@noop {} {\bibfield  {journal} {\bibinfo  {journal} {Current opinion in
  cell biology}\ }\textbf {\bibinfo {volume} {58}},\ \bibinfo {pages} {142}
  (\bibinfo {year} {2019})}\BibitemShut {NoStop}%
\bibitem [{\citenamefont {Banigan}\ \emph {et~al.}(2020)\citenamefont
  {Banigan}, \citenamefont {van~den Berg}, \citenamefont {Brand{\~a}o},
  \citenamefont {Marko},\ and\ \citenamefont {Mirny}}]{banigan2020chromosome}%
  \BibitemOpen
  \bibfield  {author} {\bibinfo {author} {\bibfnamefont {E.~J.}\ \bibnamefont
  {Banigan}}, \bibinfo {author} {\bibfnamefont {A.~A.}\ \bibnamefont {van~den
  Berg}}, \bibinfo {author} {\bibfnamefont {H.~B.}\ \bibnamefont
  {Brand{\~a}o}}, \bibinfo {author} {\bibfnamefont {J.~F.}\ \bibnamefont
  {Marko}},\ and\ \bibinfo {author} {\bibfnamefont {L.~A.}\ \bibnamefont
  {Mirny}},\ }\href@noop {} {\bibfield  {journal} {\bibinfo  {journal} {Elife}\
  }\textbf {\bibinfo {volume} {9}},\ \bibinfo {pages} {e53558} (\bibinfo {year}
  {2020})}\BibitemShut {NoStop}%
\bibitem [{\citenamefont {Alipour}\ and\ \citenamefont
  {Marko}(2012)}]{alipour2012self}%
  \BibitemOpen
  \bibfield  {author} {\bibinfo {author} {\bibfnamefont {E.}~\bibnamefont
  {Alipour}}\ and\ \bibinfo {author} {\bibfnamefont {J.~F.}\ \bibnamefont
  {Marko}},\ }\href@noop {} {\bibfield  {journal} {\bibinfo  {journal} {Nucleic
  acids research}\ }\textbf {\bibinfo {volume} {40}},\ \bibinfo {pages} {11202}
  (\bibinfo {year} {2012})}\BibitemShut {NoStop}%
\bibitem [{\citenamefont {Gibcus}\ \emph {et~al.}(2018)\citenamefont {Gibcus},
  \citenamefont {Samejima}, \citenamefont {Goloborodko}, \citenamefont
  {Samejima}, \citenamefont {Naumova}, \citenamefont {Nuebler}, \citenamefont
  {Kanemaki}, \citenamefont {Xie}, \citenamefont {Paulson}, \citenamefont
  {Earnshaw} \emph {et~al.}}]{gibcus2018pathway}%
  \BibitemOpen
  \bibfield  {author} {\bibinfo {author} {\bibfnamefont {J.~H.}\ \bibnamefont
  {Gibcus}}, \bibinfo {author} {\bibfnamefont {K.}~\bibnamefont {Samejima}},
  \bibinfo {author} {\bibfnamefont {A.}~\bibnamefont {Goloborodko}}, \bibinfo
  {author} {\bibfnamefont {I.}~\bibnamefont {Samejima}}, \bibinfo {author}
  {\bibfnamefont {N.}~\bibnamefont {Naumova}}, \bibinfo {author} {\bibfnamefont
  {J.}~\bibnamefont {Nuebler}}, \bibinfo {author} {\bibfnamefont {M.~T.}\
  \bibnamefont {Kanemaki}}, \bibinfo {author} {\bibfnamefont {L.}~\bibnamefont
  {Xie}}, \bibinfo {author} {\bibfnamefont {J.~R.}\ \bibnamefont {Paulson}},
  \bibinfo {author} {\bibfnamefont {W.~C.}\ \bibnamefont {Earnshaw}}, \emph
  {et~al.},\ }\href@noop {} {\bibfield  {journal} {\bibinfo  {journal}
  {Science}\ }\textbf {\bibinfo {volume} {359}},\ \bibinfo {pages} {eaao6135}
  (\bibinfo {year} {2018})}\BibitemShut {NoStop}%
\bibitem [{\citenamefont {Goloborodko}\ \emph
  {et~al.}(2016{\natexlab{a}})\citenamefont {Goloborodko}, \citenamefont
  {Imakaev}, \citenamefont {Marko},\ and\ \citenamefont
  {Mirny}}]{goloborodko2016compaction}%
  \BibitemOpen
  \bibfield  {author} {\bibinfo {author} {\bibfnamefont {A.}~\bibnamefont
  {Goloborodko}}, \bibinfo {author} {\bibfnamefont {M.~V.}\ \bibnamefont
  {Imakaev}}, \bibinfo {author} {\bibfnamefont {J.~F.}\ \bibnamefont {Marko}},\
  and\ \bibinfo {author} {\bibfnamefont {L.}~\bibnamefont {Mirny}},\
  }\href@noop {} {\bibfield  {journal} {\bibinfo  {journal} {Elife}\ }\textbf
  {\bibinfo {volume} {5}},\ \bibinfo {pages} {e14864} (\bibinfo {year}
  {2016}{\natexlab{a}})}\BibitemShut {NoStop}%
\bibitem [{\citenamefont {Goloborodko}\ \emph
  {et~al.}(2016{\natexlab{b}})\citenamefont {Goloborodko}, \citenamefont
  {Marko},\ and\ \citenamefont {Mirny}}]{goloborodko2016chromosome}%
  \BibitemOpen
  \bibfield  {author} {\bibinfo {author} {\bibfnamefont {A.}~\bibnamefont
  {Goloborodko}}, \bibinfo {author} {\bibfnamefont {J.~F.}\ \bibnamefont
  {Marko}},\ and\ \bibinfo {author} {\bibfnamefont {L.~A.}\ \bibnamefont
  {Mirny}},\ }\href@noop {} {\bibfield  {journal} {\bibinfo  {journal}
  {Biophysical journal}\ }\textbf {\bibinfo {volume} {110}},\ \bibinfo {pages}
  {2162} (\bibinfo {year} {2016}{\natexlab{b}})}\BibitemShut {NoStop}%
\bibitem [{\citenamefont {Hafner}\ and\ \citenamefont
  {Boettiger}(2022)}]{hafner2022spatial}%
  \BibitemOpen
  \bibfield  {author} {\bibinfo {author} {\bibfnamefont {A.}~\bibnamefont
  {Hafner}}\ and\ \bibinfo {author} {\bibfnamefont {A.}~\bibnamefont
  {Boettiger}},\ }\href@noop {} {\bibfield  {journal} {\bibinfo  {journal}
  {Nature Reviews Genetics}\ ,\ \bibinfo {pages} {1}} (\bibinfo {year}
  {2022})}\BibitemShut {NoStop}%
\bibitem [{\citenamefont {Polovnikov}\ \emph {et~al.}(2022)\citenamefont
  {Polovnikov}, \citenamefont {Belan}, \citenamefont {Imakaev}, \citenamefont
  {Brand{\~a}o},\ and\ \citenamefont {Mirny}}]{polovnikov2022fractal}%
  \BibitemOpen
  \bibfield  {author} {\bibinfo {author} {\bibfnamefont {K.}~\bibnamefont
  {Polovnikov}}, \bibinfo {author} {\bibfnamefont {S.}~\bibnamefont {Belan}},
  \bibinfo {author} {\bibfnamefont {M.}~\bibnamefont {Imakaev}}, \bibinfo
  {author} {\bibfnamefont {H.~B.}\ \bibnamefont {Brand{\~a}o}},\ and\ \bibinfo
  {author} {\bibfnamefont {L.~A.}\ \bibnamefont {Mirny}},\ }\href@noop {}
  {\bibfield  {journal} {\bibinfo  {journal} {bioRxiv}\ } (\bibinfo {year}
  {2022})}\BibitemShut {NoStop}%
\bibitem [{\citenamefont {Cattoni}\ \emph {et~al.}(2017)\citenamefont
  {Cattoni}, \citenamefont {Cardozo~Gizzi}, \citenamefont {Georgieva},
  \citenamefont {Di~Stefano}, \citenamefont {Valeri}, \citenamefont
  {Chamousset}, \citenamefont {Houbron}, \citenamefont {D{\'e}jardin},
  \citenamefont {Fiche}, \citenamefont {Gonz{\'a}lez} \emph
  {et~al.}}]{cattoni2017single}%
  \BibitemOpen
  \bibfield  {author} {\bibinfo {author} {\bibfnamefont {D.~I.}\ \bibnamefont
  {Cattoni}}, \bibinfo {author} {\bibfnamefont {A.~M.}\ \bibnamefont
  {Cardozo~Gizzi}}, \bibinfo {author} {\bibfnamefont {M.}~\bibnamefont
  {Georgieva}}, \bibinfo {author} {\bibfnamefont {M.}~\bibnamefont
  {Di~Stefano}}, \bibinfo {author} {\bibfnamefont {A.}~\bibnamefont {Valeri}},
  \bibinfo {author} {\bibfnamefont {D.}~\bibnamefont {Chamousset}}, \bibinfo
  {author} {\bibfnamefont {C.}~\bibnamefont {Houbron}}, \bibinfo {author}
  {\bibfnamefont {S.}~\bibnamefont {D{\'e}jardin}}, \bibinfo {author}
  {\bibfnamefont {J.-B.}\ \bibnamefont {Fiche}}, \bibinfo {author}
  {\bibfnamefont {I.}~\bibnamefont {Gonz{\'a}lez}}, \emph {et~al.},\
  }\href@noop {} {\bibfield  {journal} {\bibinfo  {journal} {Nature
  communications}\ }\textbf {\bibinfo {volume} {8}},\ \bibinfo {pages} {1}
  (\bibinfo {year} {2017})}\BibitemShut {NoStop}%
\bibitem [{\citenamefont {Ou}\ \emph {et~al.}(2017)\citenamefont {Ou},
  \citenamefont {Phan}, \citenamefont {Deerinck}, \citenamefont {Thor},
  \citenamefont {Ellisman},\ and\ \citenamefont {O’shea}}]{ou2017chromemt}%
  \BibitemOpen
  \bibfield  {author} {\bibinfo {author} {\bibfnamefont {H.~D.}\ \bibnamefont
  {Ou}}, \bibinfo {author} {\bibfnamefont {S.}~\bibnamefont {Phan}}, \bibinfo
  {author} {\bibfnamefont {T.~J.}\ \bibnamefont {Deerinck}}, \bibinfo {author}
  {\bibfnamefont {A.}~\bibnamefont {Thor}}, \bibinfo {author} {\bibfnamefont
  {M.~H.}\ \bibnamefont {Ellisman}},\ and\ \bibinfo {author} {\bibfnamefont
  {C.~C.}\ \bibnamefont {O’shea}},\ }\href@noop {} {\bibfield  {journal}
  {\bibinfo  {journal} {Science}\ }\textbf {\bibinfo {volume} {357}},\ \bibinfo
  {pages} {eaag0025} (\bibinfo {year} {2017})}\BibitemShut {NoStop}%
\bibitem [{\citenamefont {Bintu}\ \emph {et~al.}(2018)\citenamefont {Bintu},
  \citenamefont {Mateo}, \citenamefont {Su}, \citenamefont {Sinnott-Armstrong},
  \citenamefont {Parker}, \citenamefont {Kinrot}, \citenamefont {Yamaya},
  \citenamefont {Boettiger},\ and\ \citenamefont {Zhuang}}]{bintu2018super}%
  \BibitemOpen
  \bibfield  {author} {\bibinfo {author} {\bibfnamefont {B.}~\bibnamefont
  {Bintu}}, \bibinfo {author} {\bibfnamefont {L.~J.}\ \bibnamefont {Mateo}},
  \bibinfo {author} {\bibfnamefont {J.-H.}\ \bibnamefont {Su}}, \bibinfo
  {author} {\bibfnamefont {N.~A.}\ \bibnamefont {Sinnott-Armstrong}}, \bibinfo
  {author} {\bibfnamefont {M.}~\bibnamefont {Parker}}, \bibinfo {author}
  {\bibfnamefont {S.}~\bibnamefont {Kinrot}}, \bibinfo {author} {\bibfnamefont
  {K.}~\bibnamefont {Yamaya}}, \bibinfo {author} {\bibfnamefont {A.~N.}\
  \bibnamefont {Boettiger}},\ and\ \bibinfo {author} {\bibfnamefont
  {X.}~\bibnamefont {Zhuang}},\ }\href@noop {} {\bibfield  {journal} {\bibinfo
  {journal} {Science}\ }\textbf {\bibinfo {volume} {362}},\ \bibinfo {pages}
  {eaau1783} (\bibinfo {year} {2018})}\BibitemShut {NoStop}%
\bibitem [{\citenamefont {Nir}\ \emph {et~al.}(2018)\citenamefont {Nir},
  \citenamefont {Farabella}, \citenamefont {P{\'e}rez~Estrada}, \citenamefont
  {Ebeling}, \citenamefont {Beliveau}, \citenamefont {Sasaki}, \citenamefont
  {Lee}, \citenamefont {Nguyen}, \citenamefont {McCole}, \citenamefont
  {Chattoraj} \emph {et~al.}}]{nir2018walking}%
  \BibitemOpen
  \bibfield  {author} {\bibinfo {author} {\bibfnamefont {G.}~\bibnamefont
  {Nir}}, \bibinfo {author} {\bibfnamefont {I.}~\bibnamefont {Farabella}},
  \bibinfo {author} {\bibfnamefont {C.}~\bibnamefont {P{\'e}rez~Estrada}},
  \bibinfo {author} {\bibfnamefont {C.~G.}\ \bibnamefont {Ebeling}}, \bibinfo
  {author} {\bibfnamefont {B.~J.}\ \bibnamefont {Beliveau}}, \bibinfo {author}
  {\bibfnamefont {H.~M.}\ \bibnamefont {Sasaki}}, \bibinfo {author}
  {\bibfnamefont {S.~D.}\ \bibnamefont {Lee}}, \bibinfo {author} {\bibfnamefont
  {S.~C.}\ \bibnamefont {Nguyen}}, \bibinfo {author} {\bibfnamefont {R.~B.}\
  \bibnamefont {McCole}}, \bibinfo {author} {\bibfnamefont {S.}~\bibnamefont
  {Chattoraj}}, \emph {et~al.},\ }\href@noop {} {\bibfield  {journal} {\bibinfo
   {journal} {PLoS genetics}\ }\textbf {\bibinfo {volume} {14}},\ \bibinfo
  {pages} {e1007872} (\bibinfo {year} {2018})}\BibitemShut {NoStop}%
\bibitem [{\citenamefont {Boettiger}\ and\ \citenamefont
  {Murphy}(2020)}]{boettiger2020advances}%
  \BibitemOpen
  \bibfield  {author} {\bibinfo {author} {\bibfnamefont {A.}~\bibnamefont
  {Boettiger}}\ and\ \bibinfo {author} {\bibfnamefont {S.}~\bibnamefont
  {Murphy}},\ }\href@noop {} {\bibfield  {journal} {\bibinfo  {journal} {Trends
  in Genetics}\ }\textbf {\bibinfo {volume} {36}},\ \bibinfo {pages} {273}
  (\bibinfo {year} {2020})}\BibitemShut {NoStop}%
\bibitem [{\citenamefont {Kempfer}\ and\ \citenamefont
  {Pombo}(2020)}]{kempfer2020methods}%
  \BibitemOpen
  \bibfield  {author} {\bibinfo {author} {\bibfnamefont {R.}~\bibnamefont
  {Kempfer}}\ and\ \bibinfo {author} {\bibfnamefont {A.}~\bibnamefont
  {Pombo}},\ }\href@noop {} {\bibfield  {journal} {\bibinfo  {journal} {Nature
  Reviews Genetics}\ }\textbf {\bibinfo {volume} {21}},\ \bibinfo {pages} {207}
  (\bibinfo {year} {2020})}\BibitemShut {NoStop}%
\bibitem [{\citenamefont {Su}\ \emph {et~al.}(2020)\citenamefont {Su},
  \citenamefont {Zheng}, \citenamefont {Kinrot}, \citenamefont {Bintu},\ and\
  \citenamefont {Zhuang}}]{su2020genome}%
  \BibitemOpen
  \bibfield  {author} {\bibinfo {author} {\bibfnamefont {J.-H.}\ \bibnamefont
  {Su}}, \bibinfo {author} {\bibfnamefont {P.}~\bibnamefont {Zheng}}, \bibinfo
  {author} {\bibfnamefont {S.~S.}\ \bibnamefont {Kinrot}}, \bibinfo {author}
  {\bibfnamefont {B.}~\bibnamefont {Bintu}},\ and\ \bibinfo {author}
  {\bibfnamefont {X.}~\bibnamefont {Zhuang}},\ }\href@noop {} {\bibfield
  {journal} {\bibinfo  {journal} {Cell}\ }\textbf {\bibinfo {volume} {182}},\
  \bibinfo {pages} {1641} (\bibinfo {year} {2020})}\BibitemShut {NoStop}%
\bibitem [{\citenamefont {Liu}\ \emph {et~al.}(2020)\citenamefont {Liu},
  \citenamefont {Lu}, \citenamefont {Yang}, \citenamefont {Chen}, \citenamefont
  {Radda}, \citenamefont {Hu}, \citenamefont {Katz},\ and\ \citenamefont
  {Wang}}]{liu2020multiplexed}%
  \BibitemOpen
  \bibfield  {author} {\bibinfo {author} {\bibfnamefont {M.}~\bibnamefont
  {Liu}}, \bibinfo {author} {\bibfnamefont {Y.}~\bibnamefont {Lu}}, \bibinfo
  {author} {\bibfnamefont {B.}~\bibnamefont {Yang}}, \bibinfo {author}
  {\bibfnamefont {Y.}~\bibnamefont {Chen}}, \bibinfo {author} {\bibfnamefont
  {J.~S.}\ \bibnamefont {Radda}}, \bibinfo {author} {\bibfnamefont
  {M.}~\bibnamefont {Hu}}, \bibinfo {author} {\bibfnamefont {S.~G.}\
  \bibnamefont {Katz}},\ and\ \bibinfo {author} {\bibfnamefont
  {S.}~\bibnamefont {Wang}},\ }\href@noop {} {\bibfield  {journal} {\bibinfo
  {journal} {Nature communications}\ }\textbf {\bibinfo {volume} {11}},\
  \bibinfo {pages} {1} (\bibinfo {year} {2020})}\BibitemShut {NoStop}%
\bibitem [{\citenamefont {Xie}\ and\ \citenamefont
  {Liu}(2021)}]{xie2021single}%
  \BibitemOpen
  \bibfield  {author} {\bibinfo {author} {\bibfnamefont {L.}~\bibnamefont
  {Xie}}\ and\ \bibinfo {author} {\bibfnamefont {Z.}~\bibnamefont {Liu}},\
  }\href@noop {} {\bibfield  {journal} {\bibinfo  {journal} {Molecular Systems
  Biology}\ }\textbf {\bibinfo {volume} {17}},\ \bibinfo {pages} {e9653}
  (\bibinfo {year} {2021})}\BibitemShut {NoStop}%
\bibitem [{\citenamefont {Li}\ \emph {et~al.}(2021)\citenamefont {Li},
  \citenamefont {Eshein}, \citenamefont {Virk}, \citenamefont {Eid},
  \citenamefont {Wu}, \citenamefont {Frederick}, \citenamefont {VanDerway},
  \citenamefont {Gladstein}, \citenamefont {Huang}, \citenamefont {Shim} \emph
  {et~al.}}]{li2021nanoscale}%
  \BibitemOpen
  \bibfield  {author} {\bibinfo {author} {\bibfnamefont {Y.}~\bibnamefont
  {Li}}, \bibinfo {author} {\bibfnamefont {A.}~\bibnamefont {Eshein}}, \bibinfo
  {author} {\bibfnamefont {R.~K.}\ \bibnamefont {Virk}}, \bibinfo {author}
  {\bibfnamefont {A.}~\bibnamefont {Eid}}, \bibinfo {author} {\bibfnamefont
  {W.}~\bibnamefont {Wu}}, \bibinfo {author} {\bibfnamefont {J.}~\bibnamefont
  {Frederick}}, \bibinfo {author} {\bibfnamefont {D.}~\bibnamefont
  {VanDerway}}, \bibinfo {author} {\bibfnamefont {S.}~\bibnamefont
  {Gladstein}}, \bibinfo {author} {\bibfnamefont {K.}~\bibnamefont {Huang}},
  \bibinfo {author} {\bibfnamefont {A.~R.}\ \bibnamefont {Shim}}, \emph
  {et~al.},\ }\href@noop {} {\bibfield  {journal} {\bibinfo  {journal} {Science
  advances}\ }\textbf {\bibinfo {volume} {7}},\ \bibinfo {pages} {eabe4310}
  (\bibinfo {year} {2021})}\BibitemShut {NoStop}%
\bibitem [{\citenamefont {Gabriele}\ \emph {et~al.}(2022)\citenamefont
  {Gabriele}, \citenamefont {Brand{\~a}o}, \citenamefont {Grosse-Holz},
  \citenamefont {Jha}, \citenamefont {Dailey}, \citenamefont {Cattoglio},
  \citenamefont {Hsieh}, \citenamefont {Mirny}, \citenamefont {Zechner},\ and\
  \citenamefont {Hansen}}]{gabriele2022dynamics}%
  \BibitemOpen
  \bibfield  {author} {\bibinfo {author} {\bibfnamefont {M.}~\bibnamefont
  {Gabriele}}, \bibinfo {author} {\bibfnamefont {H.~B.}\ \bibnamefont
  {Brand{\~a}o}}, \bibinfo {author} {\bibfnamefont {S.}~\bibnamefont
  {Grosse-Holz}}, \bibinfo {author} {\bibfnamefont {A.}~\bibnamefont {Jha}},
  \bibinfo {author} {\bibfnamefont {G.~M.}\ \bibnamefont {Dailey}}, \bibinfo
  {author} {\bibfnamefont {C.}~\bibnamefont {Cattoglio}}, \bibinfo {author}
  {\bibfnamefont {T.-H.~S.}\ \bibnamefont {Hsieh}}, \bibinfo {author}
  {\bibfnamefont {L.}~\bibnamefont {Mirny}}, \bibinfo {author} {\bibfnamefont
  {C.}~\bibnamefont {Zechner}},\ and\ \bibinfo {author} {\bibfnamefont {A.~S.}\
  \bibnamefont {Hansen}},\ }\href@noop {} {\bibfield  {journal} {\bibinfo
  {journal} {Science}\ }\textbf {\bibinfo {volume} {376}},\ \bibinfo {pages}
  {496} (\bibinfo {year} {2022})}\BibitemShut {NoStop}%
\bibitem [{\citenamefont {Goloborodko}\ \emph
  {et~al.}(2016{\natexlab{c}})\citenamefont {Goloborodko}, \citenamefont
  {Marko},\ and\ \citenamefont {Mirny}}]{Goloborodko_2016}%
  \BibitemOpen
  \bibfield  {author} {\bibinfo {author} {\bibfnamefont {A.}~\bibnamefont
  {Goloborodko}}, \bibinfo {author} {\bibfnamefont {J.~F.}\ \bibnamefont
  {Marko}},\ and\ \bibinfo {author} {\bibfnamefont {L.~A.}\ \bibnamefont
  {Mirny}},\ }\href@noop {} {\bibfield  {journal} {\bibinfo  {journal}
  {Biophysical Journal}\ }\textbf {\bibinfo {volume} {110}},\ \bibinfo {pages}
  {2162} (\bibinfo {year} {2016}{\natexlab{c}})}\BibitemShut {NoStop}%
\bibitem [{\citenamefont {Banigan}\ \emph {et~al.}(2022)\citenamefont
  {Banigan}, \citenamefont {Tang}, \citenamefont {van~den Berg}, \citenamefont
  {Stocsits}, \citenamefont {Wutz}, \citenamefont {Brand{\~a}o}, \citenamefont
  {Busslinger}, \citenamefont {Peters},\ and\ \citenamefont
  {Mirny}}]{banigan2022motors}%
  \BibitemOpen
  \bibfield  {author} {\bibinfo {author} {\bibfnamefont {E.}~\bibnamefont
  {Banigan}}, \bibinfo {author} {\bibfnamefont {W.}~\bibnamefont {Tang}},
  \bibinfo {author} {\bibfnamefont {A.}~\bibnamefont {van~den Berg}}, \bibinfo
  {author} {\bibfnamefont {R.}~\bibnamefont {Stocsits}}, \bibinfo {author}
  {\bibfnamefont {G.}~\bibnamefont {Wutz}}, \bibinfo {author} {\bibfnamefont
  {H.}~\bibnamefont {Brand{\~a}o}}, \bibinfo {author} {\bibfnamefont
  {G.}~\bibnamefont {Busslinger}}, \bibinfo {author} {\bibfnamefont {J.-M.}\
  \bibnamefont {Peters}},\ and\ \bibinfo {author} {\bibfnamefont
  {L.}~\bibnamefont {Mirny}},\ }\href@noop {} {\bibfield  {journal} {\bibinfo
  {journal} {Bulletin of the American Physical Society}\ } (\bibinfo {year}
  {2022})}\BibitemShut {NoStop}%
\bibitem [{\citenamefont {Grosberg}\ and\ \citenamefont
  {Khokhlov}(1994)}]{GKh_1994}%
  \BibitemOpen
  \bibfield  {author} {\bibinfo {author} {\bibfnamefont {A.~Y.}\ \bibnamefont
  {Grosberg}}\ and\ \bibinfo {author} {\bibfnamefont {A.}~\bibnamefont
  {Khokhlov}},\ }\href@noop {} {\emph {\bibinfo {title} {Statistical Physics of
  Macromolecules}}}\ (\bibinfo  {publisher} {Woodbury, NY: AIP Press},\
  \bibinfo {year} {1994})\BibitemShut {NoStop}%
\bibitem [{\citenamefont {Belan}\ and\ \citenamefont
  {Starkov}(2022)}]{belan2022influence}%
  \BibitemOpen
  \bibfield  {author} {\bibinfo {author} {\bibfnamefont {S.}~\bibnamefont
  {Belan}}\ and\ \bibinfo {author} {\bibfnamefont {D.}~\bibnamefont
  {Starkov}},\ }\href@noop {} {\bibfield  {journal} {\bibinfo  {journal} {JETP
  Letters}\ }\textbf {\bibinfo {volume} {115}},\ \bibinfo {pages} {763}
  (\bibinfo {year} {2022})}\BibitemShut {NoStop}%
\bibitem [{\citenamefont {Darrow}\ \emph {et~al.}(2016)\citenamefont {Darrow},
  \citenamefont {Huntley}, \citenamefont {Dudchenko}, \citenamefont
  {Stamenova}, \citenamefont {Durand}, \citenamefont {Sun}, \citenamefont
  {Huang}, \citenamefont {Sanborn}, \citenamefont {Machol}, \citenamefont
  {Shamim} \emph {et~al.}}]{darrow2016deletion}%
  \BibitemOpen
  \bibfield  {author} {\bibinfo {author} {\bibfnamefont {E.~M.}\ \bibnamefont
  {Darrow}}, \bibinfo {author} {\bibfnamefont {M.~H.}\ \bibnamefont {Huntley}},
  \bibinfo {author} {\bibfnamefont {O.}~\bibnamefont {Dudchenko}}, \bibinfo
  {author} {\bibfnamefont {E.~K.}\ \bibnamefont {Stamenova}}, \bibinfo {author}
  {\bibfnamefont {N.~C.}\ \bibnamefont {Durand}}, \bibinfo {author}
  {\bibfnamefont {Z.}~\bibnamefont {Sun}}, \bibinfo {author} {\bibfnamefont
  {S.-C.}\ \bibnamefont {Huang}}, \bibinfo {author} {\bibfnamefont {A.~L.}\
  \bibnamefont {Sanborn}}, \bibinfo {author} {\bibfnamefont {I.}~\bibnamefont
  {Machol}}, \bibinfo {author} {\bibfnamefont {M.}~\bibnamefont {Shamim}},
  \emph {et~al.},\ }\href@noop {} {\bibfield  {journal} {\bibinfo  {journal}
  {Proceedings of the National Academy of Sciences}\ }\textbf {\bibinfo
  {volume} {113}},\ \bibinfo {pages} {E4504} (\bibinfo {year}
  {2016})}\BibitemShut {NoStop}%
\bibitem [{\citenamefont {Olivares-Chauvet}\ \emph {et~al.}(2016)\citenamefont
  {Olivares-Chauvet}, \citenamefont {Mukamel}, \citenamefont {Lifshitz},
  \citenamefont {Schwartzman}, \citenamefont {Elkayam}, \citenamefont
  {Lubling}, \citenamefont {Deikus}, \citenamefont {Sebra},\ and\ \citenamefont
  {Tanay}}]{olivares2016capturing}%
  \BibitemOpen
  \bibfield  {author} {\bibinfo {author} {\bibfnamefont {P.}~\bibnamefont
  {Olivares-Chauvet}}, \bibinfo {author} {\bibfnamefont {Z.}~\bibnamefont
  {Mukamel}}, \bibinfo {author} {\bibfnamefont {A.}~\bibnamefont {Lifshitz}},
  \bibinfo {author} {\bibfnamefont {O.}~\bibnamefont {Schwartzman}}, \bibinfo
  {author} {\bibfnamefont {N.~O.}\ \bibnamefont {Elkayam}}, \bibinfo {author}
  {\bibfnamefont {Y.}~\bibnamefont {Lubling}}, \bibinfo {author} {\bibfnamefont
  {G.}~\bibnamefont {Deikus}}, \bibinfo {author} {\bibfnamefont {R.~P.}\
  \bibnamefont {Sebra}},\ and\ \bibinfo {author} {\bibfnamefont
  {A.}~\bibnamefont {Tanay}},\ }\href@noop {} {\bibfield  {journal} {\bibinfo
  {journal} {Nature}\ }\textbf {\bibinfo {volume} {540}},\ \bibinfo {pages}
  {296} (\bibinfo {year} {2016})}\BibitemShut {NoStop}%
\bibitem [{\citenamefont {Beagrie}\ \emph {et~al.}(2017)\citenamefont
  {Beagrie}, \citenamefont {Scialdone}, \citenamefont {Schueler}, \citenamefont
  {Kraemer}, \citenamefont {Chotalia}, \citenamefont {Xie}, \citenamefont
  {Barbieri}, \citenamefont {de~Santiago}, \citenamefont {Lavitas},
  \citenamefont {Branco} \emph {et~al.}}]{beagrie2017complex}%
  \BibitemOpen
  \bibfield  {author} {\bibinfo {author} {\bibfnamefont {R.~A.}\ \bibnamefont
  {Beagrie}}, \bibinfo {author} {\bibfnamefont {A.}~\bibnamefont {Scialdone}},
  \bibinfo {author} {\bibfnamefont {M.}~\bibnamefont {Schueler}}, \bibinfo
  {author} {\bibfnamefont {D.~C.}\ \bibnamefont {Kraemer}}, \bibinfo {author}
  {\bibfnamefont {M.}~\bibnamefont {Chotalia}}, \bibinfo {author}
  {\bibfnamefont {S.~Q.}\ \bibnamefont {Xie}}, \bibinfo {author} {\bibfnamefont
  {M.}~\bibnamefont {Barbieri}}, \bibinfo {author} {\bibfnamefont
  {I.}~\bibnamefont {de~Santiago}}, \bibinfo {author} {\bibfnamefont {L.-M.}\
  \bibnamefont {Lavitas}}, \bibinfo {author} {\bibfnamefont {M.~R.}\
  \bibnamefont {Branco}}, \emph {et~al.},\ }\href@noop {} {\bibfield  {journal}
  {\bibinfo  {journal} {Nature}\ }\textbf {\bibinfo {volume} {543}},\ \bibinfo
  {pages} {519} (\bibinfo {year} {2017})}\BibitemShut {NoStop}%
\bibitem [{\citenamefont {Quinodoz}\ \emph {et~al.}(2018)\citenamefont
  {Quinodoz}, \citenamefont {Ollikainen}, \citenamefont {Tabak}, \citenamefont
  {Palla}, \citenamefont {Schmidt}, \citenamefont {Detmar}, \citenamefont
  {Lai}, \citenamefont {Shishkin}, \citenamefont {Bhat}, \citenamefont {Takei}
  \emph {et~al.}}]{quinodoz2018higher}%
  \BibitemOpen
  \bibfield  {author} {\bibinfo {author} {\bibfnamefont {S.~A.}\ \bibnamefont
  {Quinodoz}}, \bibinfo {author} {\bibfnamefont {N.}~\bibnamefont
  {Ollikainen}}, \bibinfo {author} {\bibfnamefont {B.}~\bibnamefont {Tabak}},
  \bibinfo {author} {\bibfnamefont {A.}~\bibnamefont {Palla}}, \bibinfo
  {author} {\bibfnamefont {J.~M.}\ \bibnamefont {Schmidt}}, \bibinfo {author}
  {\bibfnamefont {E.}~\bibnamefont {Detmar}}, \bibinfo {author} {\bibfnamefont
  {M.~M.}\ \bibnamefont {Lai}}, \bibinfo {author} {\bibfnamefont {A.~A.}\
  \bibnamefont {Shishkin}}, \bibinfo {author} {\bibfnamefont {P.}~\bibnamefont
  {Bhat}}, \bibinfo {author} {\bibfnamefont {Y.}~\bibnamefont {Takei}}, \emph
  {et~al.},\ }\href@noop {} {\bibfield  {journal} {\bibinfo  {journal} {Cell}\
  }\textbf {\bibinfo {volume} {174}},\ \bibinfo {pages} {744} (\bibinfo {year}
  {2018})}\BibitemShut {NoStop}%
\bibitem [{\citenamefont {Allahyar}\ \emph {et~al.}(2018)\citenamefont
  {Allahyar}, \citenamefont {Vermeulen}, \citenamefont {Bouwman}, \citenamefont
  {Krijger}, \citenamefont {Verstegen}, \citenamefont {Geeven}, \citenamefont
  {van Kranenburg}, \citenamefont {Pieterse}, \citenamefont {Straver},
  \citenamefont {Haarhuis} \emph {et~al.}}]{allahyar2018enhancer}%
  \BibitemOpen
  \bibfield  {author} {\bibinfo {author} {\bibfnamefont {A.}~\bibnamefont
  {Allahyar}}, \bibinfo {author} {\bibfnamefont {C.}~\bibnamefont {Vermeulen}},
  \bibinfo {author} {\bibfnamefont {B.~A.}\ \bibnamefont {Bouwman}}, \bibinfo
  {author} {\bibfnamefont {P.~H.}\ \bibnamefont {Krijger}}, \bibinfo {author}
  {\bibfnamefont {M.~J.}\ \bibnamefont {Verstegen}}, \bibinfo {author}
  {\bibfnamefont {G.}~\bibnamefont {Geeven}}, \bibinfo {author} {\bibfnamefont
  {M.}~\bibnamefont {van Kranenburg}}, \bibinfo {author} {\bibfnamefont
  {M.}~\bibnamefont {Pieterse}}, \bibinfo {author} {\bibfnamefont
  {R.}~\bibnamefont {Straver}}, \bibinfo {author} {\bibfnamefont {J.~H.}\
  \bibnamefont {Haarhuis}}, \emph {et~al.},\ }\href@noop {} {\bibfield
  {journal} {\bibinfo  {journal} {Nature genetics}\ }\textbf {\bibinfo {volume}
  {50}},\ \bibinfo {pages} {1151} (\bibinfo {year} {2018})}\BibitemShut
  {NoStop}%
\bibitem [{\citenamefont {Oudelaar}\ \emph {et~al.}(2018)\citenamefont
  {Oudelaar}, \citenamefont {Davies}, \citenamefont {Hanssen}, \citenamefont
  {Telenius}, \citenamefont {Schwessinger}, \citenamefont {Liu}, \citenamefont
  {Brown}, \citenamefont {Downes}, \citenamefont {Chiariello}, \citenamefont
  {Bianco} \emph {et~al.}}]{oudelaar2018single}%
  \BibitemOpen
  \bibfield  {author} {\bibinfo {author} {\bibfnamefont {A.~M.}\ \bibnamefont
  {Oudelaar}}, \bibinfo {author} {\bibfnamefont {J.~O.}\ \bibnamefont
  {Davies}}, \bibinfo {author} {\bibfnamefont {L.~L.}\ \bibnamefont {Hanssen}},
  \bibinfo {author} {\bibfnamefont {J.~M.}\ \bibnamefont {Telenius}}, \bibinfo
  {author} {\bibfnamefont {R.}~\bibnamefont {Schwessinger}}, \bibinfo {author}
  {\bibfnamefont {Y.}~\bibnamefont {Liu}}, \bibinfo {author} {\bibfnamefont
  {J.~M.}\ \bibnamefont {Brown}}, \bibinfo {author} {\bibfnamefont {D.~J.}\
  \bibnamefont {Downes}}, \bibinfo {author} {\bibfnamefont {A.~M.}\
  \bibnamefont {Chiariello}}, \bibinfo {author} {\bibfnamefont
  {S.}~\bibnamefont {Bianco}}, \emph {et~al.},\ }\href@noop {} {\bibfield
  {journal} {\bibinfo  {journal} {Nature genetics}\ }\textbf {\bibinfo {volume}
  {50}},\ \bibinfo {pages} {1744} (\bibinfo {year} {2018})}\BibitemShut
  {NoStop}%
\bibitem [{\citenamefont {Ulahannan}\ \emph {et~al.}(2019)\citenamefont
  {Ulahannan}, \citenamefont {Pendleton}, \citenamefont {Deshpande},
  \citenamefont {Schwenk}, \citenamefont {Behr}, \citenamefont {Dai},
  \citenamefont {Tyer}, \citenamefont {Rughani}, \citenamefont {Kudman},
  \citenamefont {Adney} \emph {et~al.}}]{ulahannan2019nanopore}%
  \BibitemOpen
  \bibfield  {author} {\bibinfo {author} {\bibfnamefont {N.}~\bibnamefont
  {Ulahannan}}, \bibinfo {author} {\bibfnamefont {M.}~\bibnamefont
  {Pendleton}}, \bibinfo {author} {\bibfnamefont {A.}~\bibnamefont
  {Deshpande}}, \bibinfo {author} {\bibfnamefont {S.}~\bibnamefont {Schwenk}},
  \bibinfo {author} {\bibfnamefont {J.~M.}\ \bibnamefont {Behr}}, \bibinfo
  {author} {\bibfnamefont {X.}~\bibnamefont {Dai}}, \bibinfo {author}
  {\bibfnamefont {C.}~\bibnamefont {Tyer}}, \bibinfo {author} {\bibfnamefont
  {P.}~\bibnamefont {Rughani}}, \bibinfo {author} {\bibfnamefont
  {S.}~\bibnamefont {Kudman}}, \bibinfo {author} {\bibfnamefont
  {E.}~\bibnamefont {Adney}}, \emph {et~al.},\ }\href@noop {} {\bibfield
  {journal} {\bibinfo  {journal} {bioRxiv}\ ,\ \bibinfo {pages} {833590}}
  (\bibinfo {year} {2019})}\BibitemShut {NoStop}%
\bibitem [{\citenamefont {Vermeulen}\ \emph {et~al.}(2020)\citenamefont
  {Vermeulen}, \citenamefont {Allahyar}, \citenamefont {Bouwman}, \citenamefont
  {Krijger}, \citenamefont {Verstegen}, \citenamefont {Geeven}, \citenamefont
  {Valdes-Quezada}, \citenamefont {Renkens}, \citenamefont {Straver},
  \citenamefont {Kloosterman} \emph {et~al.}}]{vermeulen2020multi}%
  \BibitemOpen
  \bibfield  {author} {\bibinfo {author} {\bibfnamefont {C.}~\bibnamefont
  {Vermeulen}}, \bibinfo {author} {\bibfnamefont {A.}~\bibnamefont {Allahyar}},
  \bibinfo {author} {\bibfnamefont {B.~A.}\ \bibnamefont {Bouwman}}, \bibinfo
  {author} {\bibfnamefont {P.~H.}\ \bibnamefont {Krijger}}, \bibinfo {author}
  {\bibfnamefont {M.~J.}\ \bibnamefont {Verstegen}}, \bibinfo {author}
  {\bibfnamefont {G.}~\bibnamefont {Geeven}}, \bibinfo {author} {\bibfnamefont
  {C.}~\bibnamefont {Valdes-Quezada}}, \bibinfo {author} {\bibfnamefont
  {I.}~\bibnamefont {Renkens}}, \bibinfo {author} {\bibfnamefont
  {R.}~\bibnamefont {Straver}}, \bibinfo {author} {\bibfnamefont {W.~P.}\
  \bibnamefont {Kloosterman}}, \emph {et~al.},\ }\href@noop {} {\bibfield
  {journal} {\bibinfo  {journal} {Nature Protocols}\ }\textbf {\bibinfo
  {volume} {15}},\ \bibinfo {pages} {364} (\bibinfo {year} {2020})}\BibitemShut
  {NoStop}%
\bibitem [{\citenamefont {Tavares-Cadete}\ \emph {et~al.}(2020)\citenamefont
  {Tavares-Cadete}, \citenamefont {Norouzi}, \citenamefont {Dekker},
  \citenamefont {Liu},\ and\ \citenamefont {Dekker}}]{tavares2020multi}%
  \BibitemOpen
  \bibfield  {author} {\bibinfo {author} {\bibfnamefont {F.}~\bibnamefont
  {Tavares-Cadete}}, \bibinfo {author} {\bibfnamefont {D.}~\bibnamefont
  {Norouzi}}, \bibinfo {author} {\bibfnamefont {B.}~\bibnamefont {Dekker}},
  \bibinfo {author} {\bibfnamefont {Y.}~\bibnamefont {Liu}},\ and\ \bibinfo
  {author} {\bibfnamefont {J.}~\bibnamefont {Dekker}},\ }\href@noop {}
  {\bibfield  {journal} {\bibinfo  {journal} {Nature structural \& molecular
  biology}\ }\textbf {\bibinfo {volume} {27}},\ \bibinfo {pages} {1105}
  (\bibinfo {year} {2020})}\BibitemShut {NoStop}%
\bibitem [{\citenamefont {Quinodoz}\ \emph {et~al.}(2022)\citenamefont
  {Quinodoz}, \citenamefont {Bhat}, \citenamefont {Chovanec}, \citenamefont
  {Jachowicz}, \citenamefont {Ollikainen}, \citenamefont {Detmar},
  \citenamefont {Soehalim},\ and\ \citenamefont
  {Guttman}}]{quinodoz2022sprite}%
  \BibitemOpen
  \bibfield  {author} {\bibinfo {author} {\bibfnamefont {S.~A.}\ \bibnamefont
  {Quinodoz}}, \bibinfo {author} {\bibfnamefont {P.}~\bibnamefont {Bhat}},
  \bibinfo {author} {\bibfnamefont {P.}~\bibnamefont {Chovanec}}, \bibinfo
  {author} {\bibfnamefont {J.~W.}\ \bibnamefont {Jachowicz}}, \bibinfo {author}
  {\bibfnamefont {N.}~\bibnamefont {Ollikainen}}, \bibinfo {author}
  {\bibfnamefont {E.}~\bibnamefont {Detmar}}, \bibinfo {author} {\bibfnamefont
  {E.}~\bibnamefont {Soehalim}},\ and\ \bibinfo {author} {\bibfnamefont
  {M.}~\bibnamefont {Guttman}},\ }\href@noop {} {\bibfield  {journal} {\bibinfo
   {journal} {Nature protocols}\ }\textbf {\bibinfo {volume} {17}},\ \bibinfo
  {pages} {36} (\bibinfo {year} {2022})}\BibitemShut {NoStop}%
\bibitem [{\citenamefont {Pedler}(1971)}]{Pedler_1971}%
  \BibitemOpen
  \bibfield  {author} {\bibinfo {author} {\bibfnamefont {P.}~\bibnamefont
  {Pedler}},\ }\href@noop {} {\bibfield  {journal} {\bibinfo  {journal}
  {Journal of Applied Probability}\ }\textbf {\bibinfo {volume} {8}},\ \bibinfo
  {pages} {381} (\bibinfo {year} {1971})}\BibitemShut {NoStop}%
\bibitem [{\citenamefont {Liu}\ \emph {et~al.}(2019)\citenamefont {Liu},
  \citenamefont {Kim},\ and\ \citenamefont {Hyeon}}]{liu2019heterogeneous}%
  \BibitemOpen
  \bibfield  {author} {\bibinfo {author} {\bibfnamefont {L.}~\bibnamefont
  {Liu}}, \bibinfo {author} {\bibfnamefont {M.~H.}\ \bibnamefont {Kim}},\ and\
  \bibinfo {author} {\bibfnamefont {C.}~\bibnamefont {Hyeon}},\ }\href@noop {}
  {\bibfield  {journal} {\bibinfo  {journal} {Biophysical journal}\ }\textbf
  {\bibinfo {volume} {117}},\ \bibinfo {pages} {613} (\bibinfo {year}
  {2019})}\BibitemShut {NoStop}%
\bibitem [{\citenamefont {Liu}\ \emph {et~al.}(2021)\citenamefont {Liu},
  \citenamefont {Zhang},\ and\ \citenamefont {Hyeon}}]{liu2021extracting}%
  \BibitemOpen
  \bibfield  {author} {\bibinfo {author} {\bibfnamefont {L.}~\bibnamefont
  {Liu}}, \bibinfo {author} {\bibfnamefont {B.}~\bibnamefont {Zhang}},\ and\
  \bibinfo {author} {\bibfnamefont {C.}~\bibnamefont {Hyeon}},\ }\href@noop {}
  {\bibfield  {journal} {\bibinfo  {journal} {PLoS Computational Biology}\
  }\textbf {\bibinfo {volume} {17}},\ \bibinfo {pages} {e1009669} (\bibinfo
  {year} {2021})}\BibitemShut {NoStop}%
\bibitem [{\citenamefont {Bak}\ \emph {et~al.}(2021)\citenamefont {Bak},
  \citenamefont {Kim}, \citenamefont {Liu},\ and\ \citenamefont
  {Hyeon}}]{bak2021unified}%
  \BibitemOpen
  \bibfield  {author} {\bibinfo {author} {\bibfnamefont {J.~H.}\ \bibnamefont
  {Bak}}, \bibinfo {author} {\bibfnamefont {M.~H.}\ \bibnamefont {Kim}},
  \bibinfo {author} {\bibfnamefont {L.}~\bibnamefont {Liu}},\ and\ \bibinfo
  {author} {\bibfnamefont {C.}~\bibnamefont {Hyeon}},\ }\href@noop {}
  {\bibfield  {journal} {\bibinfo  {journal} {PLoS computational biology}\
  }\textbf {\bibinfo {volume} {17}},\ \bibinfo {pages} {e1008834} (\bibinfo
  {year} {2021})}\BibitemShut {NoStop}%
\bibitem [{\citenamefont {Shi}\ and\ \citenamefont
  {Thirumalai}(2022)}]{shi2022method}%
  \BibitemOpen
  \bibfield  {author} {\bibinfo {author} {\bibfnamefont {G.}~\bibnamefont
  {Shi}}\ and\ \bibinfo {author} {\bibfnamefont {D.}~\bibnamefont
  {Thirumalai}},\ }\href@noop {} {\bibfield  {journal} {\bibinfo  {journal}
  {bioRxiv}\ } (\bibinfo {year} {2022})}\BibitemShut {NoStop}%
\bibitem [{\citenamefont {Hafner}\ \emph {et~al.}(2022)\citenamefont {Hafner},
  \citenamefont {Park}, \citenamefont {Berger}, \citenamefont {Nora},\ and\
  \citenamefont {Boettiger}}]{hafner2022beyond}%
  \BibitemOpen
  \bibfield  {author} {\bibinfo {author} {\bibfnamefont {A.}~\bibnamefont
  {Hafner}}, \bibinfo {author} {\bibfnamefont {M.}~\bibnamefont {Park}},
  \bibinfo {author} {\bibfnamefont {S.~E.}\ \bibnamefont {Berger}}, \bibinfo
  {author} {\bibfnamefont {E.}~\bibnamefont {Nora}},\ and\ \bibinfo {author}
  {\bibfnamefont {A.~N.}\ \bibnamefont {Boettiger}},\ }\href@noop {} {\bibfield
   {journal} {\bibinfo  {journal} {bioRxiv}\ } (\bibinfo {year}
  {2022})}\BibitemShut {NoStop}%
\bibitem [{\citenamefont {Mach}\ \emph {et~al.}(2022)\citenamefont {Mach},
  \citenamefont {Kos}, \citenamefont {Zhan}, \citenamefont {Cramard},
  \citenamefont {Gaudin}, \citenamefont {T{\"u}nnermann}, \citenamefont
  {Marchi}, \citenamefont {Eglinger}, \citenamefont {Zuin}, \citenamefont
  {Kryzhanovska} \emph {et~al.}}]{mach2022live}%
  \BibitemOpen
  \bibfield  {author} {\bibinfo {author} {\bibfnamefont {P.}~\bibnamefont
  {Mach}}, \bibinfo {author} {\bibfnamefont {P.~I.}\ \bibnamefont {Kos}},
  \bibinfo {author} {\bibfnamefont {Y.}~\bibnamefont {Zhan}}, \bibinfo {author}
  {\bibfnamefont {J.}~\bibnamefont {Cramard}}, \bibinfo {author} {\bibfnamefont
  {S.}~\bibnamefont {Gaudin}}, \bibinfo {author} {\bibfnamefont
  {J.}~\bibnamefont {T{\"u}nnermann}}, \bibinfo {author} {\bibfnamefont
  {E.}~\bibnamefont {Marchi}}, \bibinfo {author} {\bibfnamefont
  {J.}~\bibnamefont {Eglinger}}, \bibinfo {author} {\bibfnamefont
  {J.}~\bibnamefont {Zuin}}, \bibinfo {author} {\bibfnamefont {M.}~\bibnamefont
  {Kryzhanovska}}, \emph {et~al.},\ }\href@noop {} {\bibfield  {journal}
  {\bibinfo  {journal} {BioRxiv}\ } (\bibinfo {year} {2022})}\BibitemShut
  {NoStop}%
\bibitem [{\citenamefont {Beckwith}\ \emph {et~al.}(2022)\citenamefont
  {Beckwith}, \citenamefont {{\O}deg{\aa}rd-Fougner}, \citenamefont {Morero},
  \citenamefont {Barton}, \citenamefont {Schueder}, \citenamefont {Tang},
  \citenamefont {Alexander}, \citenamefont {Peters}, \citenamefont {Jungmann},
  \citenamefont {Birney} \emph {et~al.}}]{beckwith2022visualization}%
  \BibitemOpen
  \bibfield  {author} {\bibinfo {author} {\bibfnamefont {K.}~\bibnamefont
  {Beckwith}}, \bibinfo {author} {\bibfnamefont {{\O}.}~\bibnamefont
  {{\O}deg{\aa}rd-Fougner}}, \bibinfo {author} {\bibfnamefont {N.}~\bibnamefont
  {Morero}}, \bibinfo {author} {\bibfnamefont {C.}~\bibnamefont {Barton}},
  \bibinfo {author} {\bibfnamefont {F.}~\bibnamefont {Schueder}}, \bibinfo
  {author} {\bibfnamefont {W.}~\bibnamefont {Tang}}, \bibinfo {author}
  {\bibfnamefont {S.}~\bibnamefont {Alexander}}, \bibinfo {author}
  {\bibfnamefont {J.}~\bibnamefont {Peters}}, \bibinfo {author} {\bibfnamefont
  {R.}~\bibnamefont {Jungmann}}, \bibinfo {author} {\bibfnamefont
  {E.}~\bibnamefont {Birney}}, \emph {et~al.},\ }\href@noop {} {\bibfield
  {journal} {\bibinfo  {journal} {BioRxiv}\ ,\ \bibinfo {pages} {2021}}
  (\bibinfo {year} {2022})}\BibitemShut {NoStop}%
\bibitem [{\citenamefont {Sasaki}\ \emph {et~al.}(2022)\citenamefont {Sasaki},
  \citenamefont {Kishi}, \citenamefont {Wu}, \citenamefont {Beliveau},\ and\
  \citenamefont {Yin}}]{sasaki2022quantitative}%
  \BibitemOpen
  \bibfield  {author} {\bibinfo {author} {\bibfnamefont {H.~M.}\ \bibnamefont
  {Sasaki}}, \bibinfo {author} {\bibfnamefont {J.~Y.}\ \bibnamefont {Kishi}},
  \bibinfo {author} {\bibfnamefont {C.-t.}\ \bibnamefont {Wu}}, \bibinfo
  {author} {\bibfnamefont {B.~J.}\ \bibnamefont {Beliveau}},\ and\ \bibinfo
  {author} {\bibfnamefont {P.}~\bibnamefont {Yin}},\ }\href@noop {} {\bibfield
  {journal} {\bibinfo  {journal} {bioRxiv}\ } (\bibinfo {year}
  {2022})}\BibitemShut {NoStop}%
\bibitem [{\citenamefont {Gassler}\ \emph {et~al.}(2017)\citenamefont
  {Gassler}, \citenamefont {Brand{\~a}o}, \citenamefont {Imakaev},
  \citenamefont {Flyamer}, \citenamefont {Ladst{\"a}tter}, \citenamefont
  {Bickmore}, \citenamefont {Peters}, \citenamefont {Mirny},\ and\
  \citenamefont {Tachibana}}]{gassler2017mechanism}%
  \BibitemOpen
  \bibfield  {author} {\bibinfo {author} {\bibfnamefont {J.}~\bibnamefont
  {Gassler}}, \bibinfo {author} {\bibfnamefont {H.~B.}\ \bibnamefont
  {Brand{\~a}o}}, \bibinfo {author} {\bibfnamefont {M.}~\bibnamefont
  {Imakaev}}, \bibinfo {author} {\bibfnamefont {I.~M.}\ \bibnamefont
  {Flyamer}}, \bibinfo {author} {\bibfnamefont {S.}~\bibnamefont
  {Ladst{\"a}tter}}, \bibinfo {author} {\bibfnamefont {W.~A.}\ \bibnamefont
  {Bickmore}}, \bibinfo {author} {\bibfnamefont {J.-M.}\ \bibnamefont
  {Peters}}, \bibinfo {author} {\bibfnamefont {L.~A.}\ \bibnamefont {Mirny}},\
  and\ \bibinfo {author} {\bibfnamefont {K.}~\bibnamefont {Tachibana}},\
  }\href@noop {} {\bibfield  {journal} {\bibinfo  {journal} {The EMBO journal}\
  }\textbf {\bibinfo {volume} {36}},\ \bibinfo {pages} {3600} (\bibinfo {year}
  {2017})}\BibitemShut {NoStop}%
\bibitem [{\citenamefont {De~Gennes}(1979)}]{DeGennes_1979}%
  \BibitemOpen
  \bibfield  {author} {\bibinfo {author} {\bibfnamefont {P.-G.}\ \bibnamefont
  {De~Gennes}},\ }\href@noop {} {\emph {\bibinfo {title} {Scaling concepts in
  polymer physics}}}\ (\bibinfo  {publisher} {Cornell University Press},\
  \bibinfo {year} {1979})\BibitemShut {NoStop}%
\bibitem [{\citenamefont {Abramowitz}\ \emph {et~al.}(1988)\citenamefont
  {Abramowitz}, \citenamefont {Stegun},\ and\ \citenamefont
  {Romer}}]{abramowitz1988handbook}%
  \BibitemOpen
  \bibfield  {author} {\bibinfo {author} {\bibfnamefont {M.}~\bibnamefont
  {Abramowitz}}, \bibinfo {author} {\bibfnamefont {I.~A.}\ \bibnamefont
  {Stegun}},\ and\ \bibinfo {author} {\bibfnamefont {R.~H.}\ \bibnamefont
  {Romer}},\ }\href@noop {} {\bibinfo {title} {Handbook of mathematical
  functions with formulas, graphs, and mathematical tables}} (\bibinfo {year}
  {1988})\BibitemShut {NoStop}%
\end{thebibliography}%

\end{document}